\documentclass[manuscript]{aastex}
\usepackage{epstopdf}
\usepackage{graphicx}

\def\Msun{M_\odot}

\slugcomment{Submitted to AJ}

\shorttitle{GK Per with HST}
\shortauthors{Shara et al.}

\begin{document}

\title{GK Per (Nova Persei 1901): HST Imagery and Spectroscopy of the Ejecta, and First Spectrum of the Jet-Like Feature}


\author{Michael~M.~Shara\altaffilmark{1,4}, David~Zurek\altaffilmark{1,4}, Orsola De Marco\altaffilmark{2}, Trisha~Mizusawa\altaffilmark{1}, Robert Williams\altaffilmark{3} and Mario Livio\altaffilmark{3}}


\altaffiltext{1}{Department of Astrophysics, American Museum of Natural
History, Central Park West and 79th street, New York, NY 10024-5192}
\altaffiltext{2}{Department of Physics, Macquarie University, Sydney, Australia}
\altaffiltext{3}{Space Telescope Science Institute, 3700 San Martin Drive, Baltimore MD 21218}
\altaffiltext{4} {Visiting Astronomer, Kitt Peak National Observatory, National Optical Astronomy Observatory, which is operated by the Association of Universities for Research in Astronomy (AURA) under cooperative agreement with the National Science Foundation. }

\begin{abstract}

We have imaged the ejecta of GK Persei (Nova Persei 1901 A.D.) with the Hubble Space Telescope (HST), whose 0.1 arcsec resolution reveals hundreds of cometary-like structures with long axes aligned towards GK Per.  One or both ends of the structures often show a brightness enhancement relative to the structures' middle sections, but there is no simple regularity to their morphologies (in contrast with, for example, the Helix nebula). Some of structures' morphologies suggest the presence of slow-moving or stationary material with which the ejecta is colliding, while others suggest shaping from a wind emanating from GK Per itself. The most detailed expansion map of any classical nova's ejecta was created by comparing HST images taken in successive years. Wide Field and Planetary Camera 2 narrowband images and Space Telescope Imaging Spectrograph spectra demonstrate that the physical conditions in this nova's ejecta vary strongly on spatial scales much smaller than those of the ejecta. Directly measuring accurate densities and compositions, and hence masses of this and other nova shells, will demand data at least as resolved spatially as those presented here. The filling factor the ejecta is 1 \% or less, and the nova ejecta mass must be less than $10^{-4} \Msun$. A modest fraction of the emission nebulosities vary in brightness by up to a factor of two on timescales of one year. Finally, we present the deepest images yet obtained of a jet-like feature outside the main body of GK Per nebulosity, and the first spectrum of that feature. Dominated by strong, narrow emission lines of [NII], [OII], [OIII], and [SII], this feature is probably a shock due to ejected material running into stationary ISM, slowly moving ejecta from a previous nova episode, or circum-binary matter present before 1901. An upper limit to the mass of the jet is of order a few times $10^{-6} \Msun$.  If the jet mass is close to this limit then the jet might be an important, or even dominant mass sink from the binary system. The jet's faintness suggests that similar features could easily have been missed in other cataclysmic binaries.

\end{abstract}

\keywords{binaries: novae, cataclysmic variables --- binaries: close --- novae---stars:outflows}

\section{Introduction}

The ejecta of old novae are particularly interesting because they are probes of both the thermonuclear processes and eruptions that give rise to classical novae, and of the interstellar matter surrounding cataclysmic variables. One of the oldest, largest and brightest nova shells is that of GK Persei, the very fast nova of 1901. The presence of very strong lines of Neon \citep{cpg57} and Magnesium \citep{eva92} suggest that GK Per was an ONeMg nova. GK Per reached visual magnitude 0.2 shortly after it erupted on 22 February, 1901. Possibly the first observations of its color and spectrum were conducted by \citet{fro01} at the Dartmouth Observatory, who wrote: ``The brilliant object attracted my attention at eleven o'clock on the evening of February 22nd, before the receipt of the announcement of its discovery by Dr. Anderson. It was at that time to my eye brighter than a standard first-magnitude star, and showed a distinct yellowish color, recalling to my mind the shade of Nova Aurigae...  Although the spectroscope employed does not permit micrometer settings to be made, the identification would seem to be sufficiently exact of the hydrogen lines H$\alpha$ and H$\beta$, the sodium lines at D, the magnesium group b (in whole or part),...all these being represented by bright and dark components. Numerous other lines were seen that cannot yet be identified."

Soon afterwards \citet{per01}, \citet{per02} and \citet{rit02} published striking images of the nebulosity surrounding GK Per. \citet{kap01} correctly suggested that that a light echo was being observed. \citet{cou39} explained that forward scattering of light by dust sheets along the line of sight to the nova could explain the otherwise superluminal velocities. Narrowband images in the lines of hydrogen, oxygen and especially [NII] show an elongated, boxy shell of expanding, ionized gas almost 2 arcmin in size surrounding the central star of GK Per  (\citet{anu93} and \citet{law95}). The narrowband imaging of \citet{sla95} shows in detail the differences in morphology between the ejecta imaged in the light of different ions. \citet{sla95} note that the [OIII] images of the northeast and southwest quadrants suggest different physical conditions, or that the ejecta in the southeast region have ``burst out" from a high- to a low-density environment. This also appears to be the case in the eastern ``hole" seen in all the narrowband images of all ions.

Very Large Array detections of GK Per showed it to be a non-thermal, polarized radio source \citep{rey84} displaying shocked circumstellar or interstellar material \citep{bie95}. IRAS images (especially those at 100 microns) examined by \citet{bod87} led to the discovery of an extended bipolar nebula around GK Per, leading \citet{bod87} to suggest that GK Per was embedded in an ancient planetary nebula (see also \citet{dou96}).  \citet{twe95} pointed out that the white dwarf in GK Per must be at least $100\times$ older and $5-10\times$ colder than planetary nebula white dwarfs to have undergone a nova explosion. The extended nebula was  also detected in HI \citep{sea89} and CO \citep{hes89} and \citep{sco94}. While some of the IRAS emission may be due to interstellar cirrus \citep{hes89}, the conclusion that ejecta from GK Per is colliding with slow-moving or stationary exterior matter seems firm.  International Ultraviolet Explorer spectra of the GK Per shell \citep{eva92} suggest the presence of abundance gradients, in which excitation is likely to be due to shocks rather than by the weak ultraviolet emission of the stellar remnant.

GK Per was the first classical nova shell detected in X-rays \citep{bal99}. ROSAT High Resolution Imager data showed that the X-ray nebula is clumpy, with an X-ray temperature of $2\times10^{6}$~K, and that it has an elliptical shape like that seen in optical narrowband images. The clumps were ascribed to fragmentation and condensation in a reverse shock zone. Chandra ACIS spectra \citep{bal05} show that the ejecta has roughly solar composition except for significant overabundances, relative to solar, of Nitrogen and Neon (which again suggests that GK Per was an ONeMg nova). Emission measures suggest electron densities ranging from 0.6 to 11 cm$^{-2}$, but only for assumed filling factors of unity. The optical narrowband images showing highly clumped matter suggest densities 10 to 100 times higher. The cooling seen in the forward shock in the X-ray images is also detected in the optical narrowband [NII] images.

GK Per is one of a few old novae that show regular dwarf nova eruptions, which have been described in detail by \citet{bia86}; see also \citet{eva09}. Winds up to few thousand km/s accompany dwarf nova eruptions \citep{fro02}; this fact will be noted when we describe high resolution imagery of GK Per's ejecta.

A remarkable feature is seen in very deep narrowband optical images of GK Per: a jet-like feature in the northeast quadrant, first noted by \citet{anu93}, and first described by \citet{bod04}. The jet appears to curve and spread as it stretches outwards, and to have its origin at the center of the ejecta. Its origin and connection with the faint, extended nebulosity around GK Per are uncertain,
but its geometry suggests an intimate connection.

All of the intriguing observations noted above are limited in angular resolution to about 1 arcsec. At 470 parsec, GK Per is closer and better resolved than all old nova shells except that of Z Cam \citep{sha07}. The "smooth" ejected shell of the recurrent nova T Pyxidis detected in groundbased images \citep{sha89} was resolved into thousands of knots with Hubble Space Telescope (HST) images \citep{sha97}, yielding significant insight into the longterm evolution of this object \citep{sps10}. GK Per is several times closer to the Sun than T Pyx, so that HST images of the ejecta and nebulosities around GK Per might yield even more detail, and hopefully, a better understanding of their origin and evolution. For this reason we applied for, and were granted HST time to obtain images and spectra of GK Per. We also obtained deeper ground-based images of the jet-like feature, and its first spectrum. 
 
In section 2 we describe the HST and ground-based datasets. HST narrowband images of GK Per are shown in section 3, and an expansion map derived from some of these images is presented in section 4. Spectra of six of the cometary structures that we have discovered are shown and analyzed in section 5. A very deep, narrowband image of the jet-like structure is shown in section 6, as is its first spectrum. We summarize our results in section 7.

\section{The Data}

We observed GK Per under the auspices of HST proposals  6060, 6965 and 7768. Narrowband images in the lines of NeV, [OII],  HeII, H$\beta$, [OIII], H$\alpha$ and [NII] were obtained with the Wide Field and Planetary Camera 2 (WFPC2) through the F343N, F375N, F469N, F487N, F502N, F656N and F658N filters, respectively. We also obtained broadband comparison images through the F450W, F555W and F675W filters. Followup long-slit spectroscopy was obtained with the Hubble Space Telescope Imaging Spectrograph (STIS) on 29 January, 1999, with the G430L (3000-5700~\AA ) and G750L (5200-10\,000~\AA ) gratings (R$\sim$2000), through the 52$\times$2~arcsec$^2$ slit. Details of dates and exposure times for our HST imaging and spectroscopy are listed in Table 1. Ground-based [NII] images of GK Per were obtained with the Mayall 4 meter telescope and  CCD Mosaic imager at Kitt Peak National Observatory on 06 and 07 February, 2010. Spectra of the jet-like feature (described in section 6) were obtained on 10 February, 2010 with the RC spectrograph of the Mayall telescope. These too are listed in Table 1.

\section{HST images of GK Per}

The HST images for each filter obtained in October 1995 and January 1997 (see Table 1) were combined using multidrizzle and associated software in IRAF/STSDAS to remove cosmic rays and to mosaic the 4 chips into a single image. The images in the lines of NeV, [OII],  HeII, and H$\beta$ showed little or no detail; emission in these passbands is much weaker than in [OIII], H$\alpha$ and [NII] .

In Figure 1 we show the combined [NII] image of GK Per. In Figure 2 we show the combined H$\alpha$ image of GK Per, while Figure 3 is the combined [OIII] image. Figure 4 is a color image where red is [NII], green is H$\alpha$ and blue is [OIII]. The emission asymmetries already noted by other authors (\citet{anu93} and \citet{sla95}) are seen in Figure 4. In particular, the enhancement of [OIII] emission in the northern half of the shell relative to the southern ejecta is apparent. [NII] emission dominates in the south.

In Figure 5 we show a close-up of a section of the [NII] image. Ground-based images of GK Per have shown that it is very inhomogeneous, composed of hundreds of knots. A key finding of this paper is the detection of almost one thousand cometary-like structures, evident in Figure 5. These features are almost all aligned with their long axes pointing at the central star. This is the first detection of such features in the ejecta of classical novae. The many hundreds of knots seen in HST images of T Pyxidis might, in fact, be similar structures. Because T Pyx is at least seven times farther away than GK Per (the closest classical nova of the 20th century) we are probably not resolving its knots. In Figure 6 we show a mosaic closeup (displaying the three filters in which the knots are visible) of several of the cometary features.

The cometary structures seen in Figures 5 and 6 range up to 12 arcsec or so in length, corresponding to linear sizes of about $10^{17}$ cm: of order a tenth the dimensions of the ejecta. The widths of these same features are barely (if at all) resolved, corresponding to dimensions of $10^{15}$ cm or less...the size of the solar system. The broadest and brightest parts of the cometary structures are sometimes closest to the central star, and sometimes furthest from it. Cometary-shaped features are seen in planetary nebulae (especially the Helix nebula, NGC 7293 -  \citet{mea10}). In contrast with the cometary features seen in GK Per, the Helix nebula's cometary features, shaped by an outflowing wind, invariably  display their narrowest widths closest to the central star.   

In planetary nebulae the knots are highly radially symmetric, each containing bright cusps (which are local ionization fronts) and tails. All extend away from the planetary central star in a radial direction. The cometary features in planetary nebulae are explained as accelerating ionized flows of ablated material driven by the previous, mildly supersonic, asymptotic giant branch wind from the central star \citep{mea10}. As we will show below, the spectrum of the accretion disk of GK Per is dominated by Balmer lines, (rather than, say, HeII lines), and there is no HeII emission seen in the inner parts of the GK Per nebula in our narrowband HeII images. Thus the accreting white dwarf is unlikely to be a sufficiently luminous source of ionizing radiation to photoionize the ejecta. The outer, narrower parts of the comet-like features often terminate in a bright knot, suggestive of a shock interaction of the outflowing ejecta with a stationary or slower moving medium. The observational case for large amounts of circum-binary matter surrounding, and interacting with the ejecta of novae is given in \citet{wil10}. The morphologies we see in the outer parts of some of the GK Per comet-like features are consistent with blobs of ejecta running into a pre-existing ambient medium. We have already noted the detection of large regions of faint surface brightness material outside the 2 arcmin diameter ejecta of GK Per, and will show further evidence in section  6 below. 

\section{Expansion and Variability of the GK Per ejecta}

We used contour plots of intensity to identify the brightest peaks in each knot in each epoch of the [NII] images. In all we identified and matched 937 knots.  We determined the centroid of light of each knot positions using the IRAF task centroid. We then calculated the proper motion of the centroid of light of each knot, and did the same for a sample of 25 stars in the GK Per image. In Figure 7a we show the map of the motion of the 937 knots. In Figure 7b we show, to the same scale, the near absence of motion detected in the 25 stars. 

In Figure 8 we show the amplitude of the knots' proper motions as a function of radius. A pure linear relationship (shown as a solid line in the figure) between proper motion and radial distance from the nova is a fairly good fit the data. The vertical scatter in the plotted data is several times larger than the error bar shown in the upper left of the figure. It is due to the asymmetry in the nova ejection velocity field seen in Figure 7a. If the ejection of material from the white dwarf in the GK Per binary was nearly instantaneous, spherically symmetric, and into a vacuum then we would expect a thin-shell spherical nebula with a smoothly varying surface-brightness distribution. From the morphology of Figures 1 through 6 we see that neither the ``spherically symmetric" nor the ``surrounding vacuum" approximations can be valid. There is clearly an azimuthal dependence on the ejection velocity - a phenomenon often observed in nova shells which are characteristically prolate with equatorial bands and polar caps \citet{for88}. The behavior of the ejecta, subject to deceleration, has been modeled by \citet{sea89}. \citet{due87} has measured the deceleration of the GK Per ejecta, finding a mean knot angular velocity of about 0.3 arcsec/yr, and showing that the time for individual knots to decrease in speed by a factor of two was 58 years. The largest angular velocity measured for any knot was $0.403$ arcsec/yr while the largest angular velocity perpendicular to that direction was $0.330$ arcsec/yr. These numbers are consistent with our own measurements.

In Figure 9 we show the comparison between the brightnesses of knots in the [NII] images measured in October 1995 and January 1997. Most of the knots are constant in brightness, with brightness changes smaller than the measurement errors. Remarkably, though, there are knots which exhibit striking brightness differences - up to a full magnitude. (Knots on the left side of Figure 10 were brighter in January 1997 while the knots to the right of most points were brighter in October 1995.)  A similar effect, on a similar timescale, has been seen in the knots of the recurrent nova T Pyxidis \citep{sha97}. A few of the most striking changes in knots' brightnesses are shown in the mosaic that is Figure 10.

\section{Spectra}  

STIS/CCD long-slit spectroscopy of the GK Per nova ejecta was obtained on 29 January, 1999 with the G430L (3000-5700~\AA ) and G750L (5200-10\,000~\AA ) gratings (R$\sim$2000), through the 52$\times$2~arcsec$^2$ slit. The slit was positioned on part of the outer rim of the shell structure, and included several bright knots (as can be seen in Fig 1, where the slit position is presented superimposed on an [N~{\sc ii}] narrow-band [F658N] image of the ejecta). The resulting spectra (presented in Figure 11)  consist of very broad emission lines, but no continuum, as is expected of moderate-density nebular gas. Their complex structure on the 2-dimensional CCD image is due to the varying knot brightness (in the spatial direction) and the instrumental wavelength broadening expected for the 2-arcsec-wide slit (in the spectral direction). 

On-the-fly calibration was  used for the data. The six brightest knots in Figure 11 (a) produced spectra with sufficiently high S/N to use for analysis. The most prominent emission lines are the [O~{\sc ii}] $\lambda \lambda$3727,29 doublet (blended together), the [O~{\sc iii}] $\lambda$$\lambda$4959,5007 doublet (semi-blended) and the [N~{\sc ii}] $\lambda$$\lambda$6543,6584-H$\alpha$ complex. Knot 4 also displays H$\beta$, while knots 4 and 5 show evidence of [O~{\sc ii}] $\lambda \lambda$7320,30. One additional emission line is observed in the spectra of knots 2, 3, 4 and 5, while a second additional emission line can clearly be seen in the spectrum of knot 4. The centers of the two emission lines are measured to be at 3901~\AA\ and 3995~\AA , respectively (using knot 4, we measured the central wavelengths of the emission lines by determining the middle of their full-width-at-zero-intensity). Since H$\gamma$ is not observed in any of the knots, we exclude that these unidentified lines belong to higher members of the Balmer series. The two lines are almost certainly He~{\sc i} $\lambda$3889 and [Ne~{\sc iii}] $\lambda$3968, which are also observed in other nova shells such as Nova Cygni 1978 (Stickland et al. 1981). The wavelength shifts of these two lines (925 and 1965~km~s$^{-1}$) are consistent with those of other lines such as H$\beta$ (1600~km~s$^{-1}$), given the uncertainties due to the instrumental shifts expected for the very wide 2-arcsec slit. (The 2-arcsec-wide slit covers in fact $\sim$40 pixels in the spectral direction. This means that if a knot spans the full 40 pixels in the slit, its emission lines will be broadened by 54~\AA\ or 97~\AA , for the G430L and G750L gratings, respectively. If the knot does not span all of the 40 pixels of the slit width, the emission lines will be less broadened, but their centers will be shifted. The line brightness variations across a knot additionally contribute to its appearing more or less shifted.)

We extracted six spectra, corresponding to the brightest ejecta knots (marked 1 to 6 in Figure 11 [a,b and c]), by using 16-pixel-wide apertures. These apertures were centered, in the spatial direction, by using the [N~{\sc ii}]-H$\alpha$ line's peak intensity (marked in Figure 11 [c]). The background was subtracted, before extracting the spectra, by taking an average of the counts on either side of the emission lines in the spatial direction. The sample regions were placed manually. Some hot/warm pixels, which were not eliminated by the on-the-fly calibration, were removed by hand by interpolation. Emission line fluxes were measured by integrating above the zero level, as well as by fitting a Gaussian curve to the the emission line. Even when this did not represent the shape of the line, this method smooths the noise spikes and allows one to estimate the uncertainty. Fluxes are accurate to $\sim$15-20\%. The reddening to GK Per cannot be determined from our data using emission lines since not enough de-blended Balmer lines exist in our spectra. Other line pairs with the same upper level and a sufficient difference in their central wavelength are not available. Based on a distance of 470~pc \citep{war87} and \citep{due81}, and the nova's Galactic coordinates (150.96, -10.10), we calculated a reddening of $E(B-V)=0.32$~mag, using the dust prescription of Schlegel et al. (1998). 

From Table~2, it is clear that physical conditions and/or chemical abundances vary from knot to knot, since the de-reddened flux ratios (Table~3) are not consistent from knot to knot.  The brightest two knots (knots 4 and 5) yield enough emission lines for an electron  density diagnostic (the [O~{\sc ii}]ratio $\lambda \lambda$3727,29/$\lambda \lambda$7320,30 is sensitive to the electron density with only a small electron temperature sensitivity). From the ratios reported in Table~3, and a 5-level atom calculation (implemented in the EQUIB code written by I.D. Howarth and S. Adams (PhD thesis 1983)  and \citep{dem97}), the electron density is measured to be (3.5$\pm$2.0)$\times$10$^{3}$~cm$^{-3}$. Knots 4 and 5 have the highest [N~{\sc ii}]+H$\alpha$/[O~{\sc iii}] ratio by far (Table~3), an indication of (a) high N/O ratio, (b) low O/H ratio and/or (c) low ionization. The correspondence of the [O~{\sc ii}]/[O~{\sc iii}] ratios (an ionization diagnostic) to the [N~{\sc ii}]-H$\alpha$/[O~{\sc iii}] indicates that the reason for the spread is likely to be due to a range in ionization and electron density of the knots, rather than very different abundances. Knots 4 and 5 are likely to have the highest density and hence the lowest ionization (since recombination is a function of electron density).

It is clear from Figure 5 is that the volumetric filling factor of the GK Per ejecta is of order one percent or less. Assuming the same density throughout each cometary structure as that deduced for the two brightest knots above, a distance of 470 pc, a 1\% filling factor and and a mean shell radius of 40 arcsec yields an upper limit on the visible ejecta of roughly $2.6 \times 10^{-4} \Msun$. The large deceleration of the ejecta measured by \citet{due87} demonstrates that the majority of the visible ejecta mass must have been "swept-up" during the past century, so that the nova itself ejected less than $10^{-4} \Msun$.

\section{The "JET"}

A jet-like feature in the northeast quadrant, first noted by \citet{anu93}, and first described by \citet{bod04}, appears to curve and spread as it stretches outwards, and to have its origin at the center of the ejecta. A deep narrowband exposure (see Table 1) of the GK Per ejecta is shown in Figure 12a, extending 2 arcmin to the NW of the central star. This corresponds to 0.27 parsec at the 470 parsec distance of GK Per. The "jet" terminates at, and blends into an arc of very faint surface brightness nebulosity, also visible in Figure 12a. An equally faint band of nebulosity is seen to the SW of GK Per, stretching parallel to the band to the NE. This is clearly seen in  our stretched (and negative) net  H$\alpha$ image which is Figure 12b. The origin of the jet, and why it is apparently one-sided, is unknown. If it is caused by particles moving at the same speed as the highly visible ejecta then it must have begun forming at least a century before the 1901 eruption. The jet is not seen interior to the bright ejecta, but this may not be surprising as the space surrounding the old nova has almost certainly been ``snowplowed". This is expected, on the basis of the measured deceleration of GK Per's ejecta \citep{due87}. The curvature in the jet, especially if extended back to GK Per, suggests that it might be a rotating feature; this could connect the one-sided jet to the nebulosity to the SE. Alternately, the entire GK Per system (including the ejecta) may be moving towards the NW, and the ``wind" of a nearly stationary ISM could be sweeping the jet "backwards". The somewhat flattened morphology on the NW side of the boxy ejecta, as well as the systematically smaller proper motions to the W and NW relative to those moving towards the E and SE in Figure 7a are consistent with this interpretation. However, the lack of an enhanced leading shock edge on the NW side of the jet argues against this latter idea.

The spectrum of GK Per and its accretion disk are shown in Figure 13a, while that of the jet is shown in Figure 13b. The relative emission line strengths in the jet spectrum are listed in Table 4. [NII] dominates H$\alpha$,
with strong lines of [OIII] and [OII] also in evidence. While our resolution and S/N of the jet lines is insufficient to place any limits on its velocity, the presence of H$\alpha$ and lack of detectable HeI or HeII lines in the jet suggests that the plasma temperature cannot be very different from $10^{4}$ Kelvins. The [S~{\sc ii}] $\lambda \lambda$6716,6731 doublet is an excellent density diagnostic, and reasonably well resolved in our data. The [S~{\sc ii}] $\lambda \lambda$6716,6731 intensity ratio was measured with the IRAF deblending function as 1.40 (see also Figure 13c), corresponding (at a plasma temperature of roughly $10^{4}$ Kelvins ) to a density of 20-40 ~cm$^{-3}$ \citep{ost89}. This is about an order of magnitude less than the density reported for knots 4 and 5 in the previous section, and allows us to put limits on the mass of the jet. 

We assume that the filling factor in the jet is unity, as there is no significant clumping or brightness inhomogeneity seen along its length in Figure 12a. The length of the jet-like feature is about 1 arcmin, corresponding to $4.2\times10^{17}$~cm , while its width is roughly one-tenth that size. If we assume that the depth of the jet is comparable to its width, and that the jet has a uniform density of 20-40 ~cm$^{-3}$, we find a mass of order $5-10\times10^{-6} \Msun$. If this is the jet mass we point out that this represents a significant fraction of the mass ejected during the nova explosion. Of course, we have no way of knowing if the jet's depth is comparable to its width, so the mass just estimated should be regarded, for now, as an upper limit. If the jet is a persistent feature of the GK Per system then the mass loss it implies will play an important role in the evolution of the binary. CVs are the only systems containing an accreting compact object in which highly collimated jets have not been unambiguously identified so far (see \citet{liv00} for a review). However, the long orbital period of GK Per (almost 2 days) puts it in the same class as the supersoft x-ray sources (in which collimated jets have been observed), in terms of the ratio of outer/inner accretion disk radius. Since this may be a crucial parameter for jet collimation (e.g. \citet{spr96} and \citet{liv00}), it is possible that the jet-like feature indeed represents a collimated jet. If jets are common but very faint and so-far undetected features of old novae, or of other types of cataclysmic variables, their effects will be equally important in understanding these systems' longterm evolution.

\subsection{Summary and Conclusions}

The highest spatial resolution narrowband images ever obtained of the ejecta of an old nova are presented. They demonstrate the presence of a thousand filamentary ("comet-like") structures, including some as small as 50 Astronomical Units in size. The structures' long axes are radially aligned in the direction of GK Per. Physical conditions including ionization and electron density are varying strongly across the resolved filaments, helping place an upper limit of  $10^{-4} \Msun$ on the mass ejected by the nova in 1901. A density of $(3.5\pm2.0)\times10^{3}$~cm$^{-3}$  is measured in two filaments. Some of the filamentary structures are largest and brightest in the direction of GK Per, while others show the opposite behavior. We suggest that these features may be sculpted both by a wind flowing from the old nova and by interaction with material that surrounded GK Per before it erupted.  A detailed expansion map of the GK Per nebulosity shows a remarkably smooth variation in ejecta velocity with azimuth. A few percent of the knots are seen to vary significantly in brightness on a timescale of one year. The remarkable jet-like feature extending over 0.27 parsecs from GK Per is suggestive of the shock interaction of a collimated flow with the ISM. The jet is estimated to contain a mass with an upper limit of $5-10\times10^{-6} \Msun$.The origin of the jet -- if indeed it is a jet -- is most probably the center of the CV accretion disk. Its collimation mechanism remains uncertain, although it is probably similar to that operating in supersoft systems. If it is a persistent feature then the mass loss it signals may have a profound influence on the evolution of GK Per. The jet's faintness suggests that similar features could easily have been missed in other cataclysmic binaries.



\acknowledgments

This research is based on NASA/ESA Hubble Space Telescope observations obtained at the Space Telescope Science Institute, which is operated by the Association of Universities for Research in Astronomy Inc. under NASA contract NAS5-26555.


\clearpage

\begin{figure}
\figurenum{1}
\epsscale{1.0}
\plotone{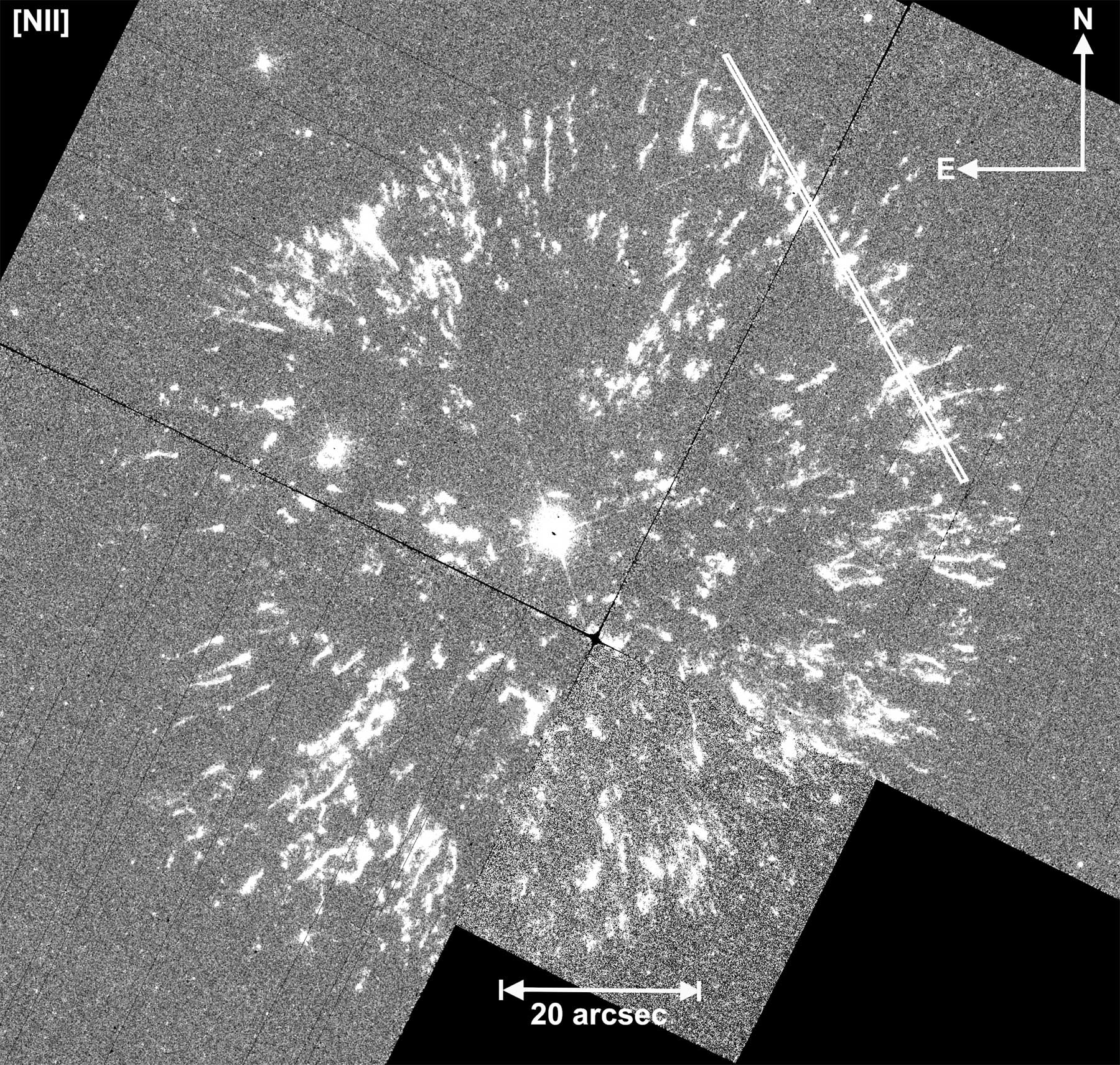}
\caption{HST/WFPC2 summed image in F658N [NII] filter of the ejecta of the old nova GK Per. The position of the slit used in STIS/CCD long-slit spectroscopy of the GK Per nova ejecta, placed on part of the outer rim of the shell structure, and including several bright knots seen in the [NII] exposures, is also shown.}
\end{figure}

\clearpage

\begin{figure}
\figurenum{2}
\epsscale{1.0}
\plotone{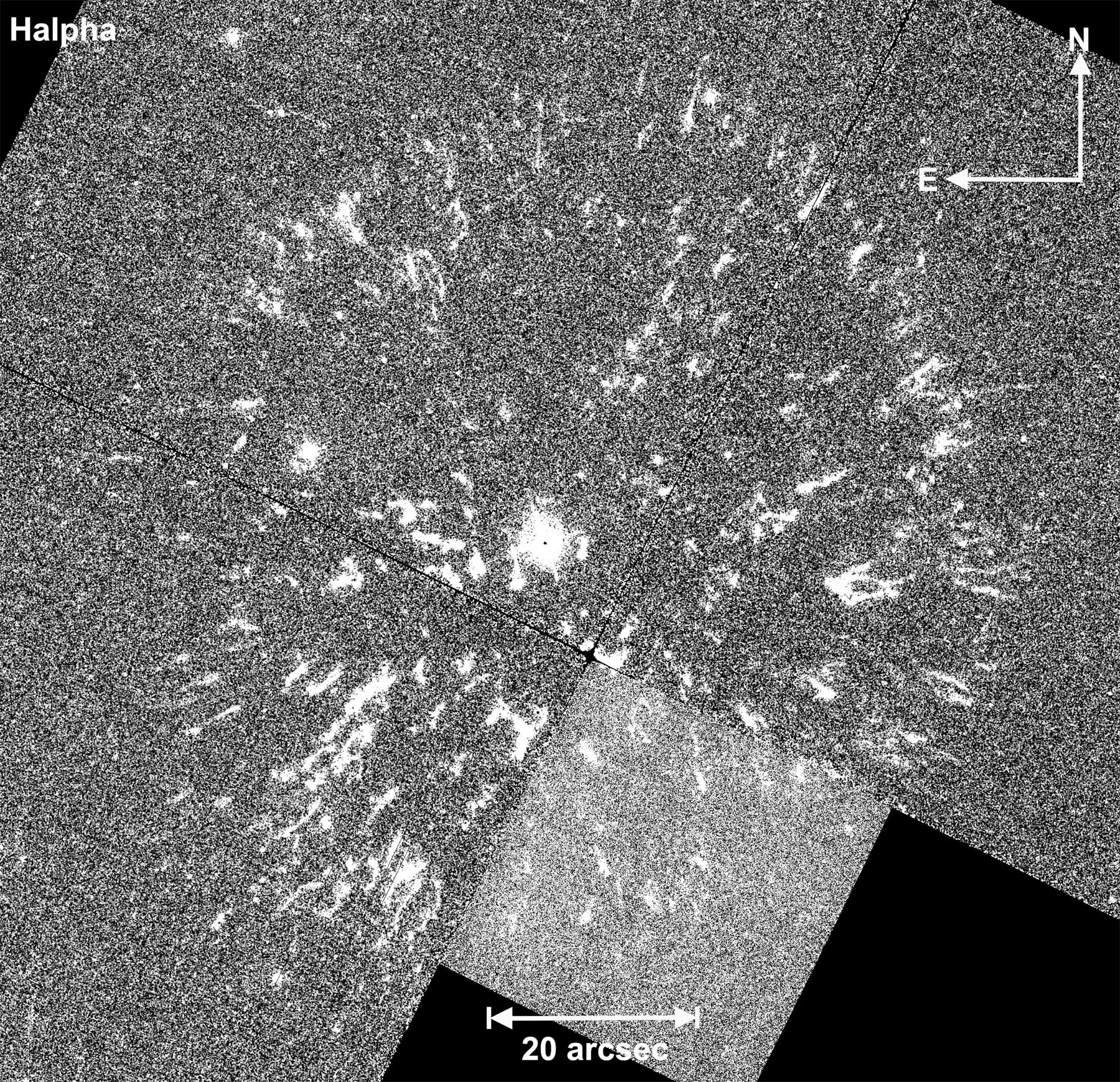}
\caption{HST/WFPC2 summed image in F656N H$\alpha$ filter of the ejecta of the old nova GK Per}
\end{figure}

\clearpage

\begin{figure}
\figurenum{3}
\epsscale{1.0}
\plotone{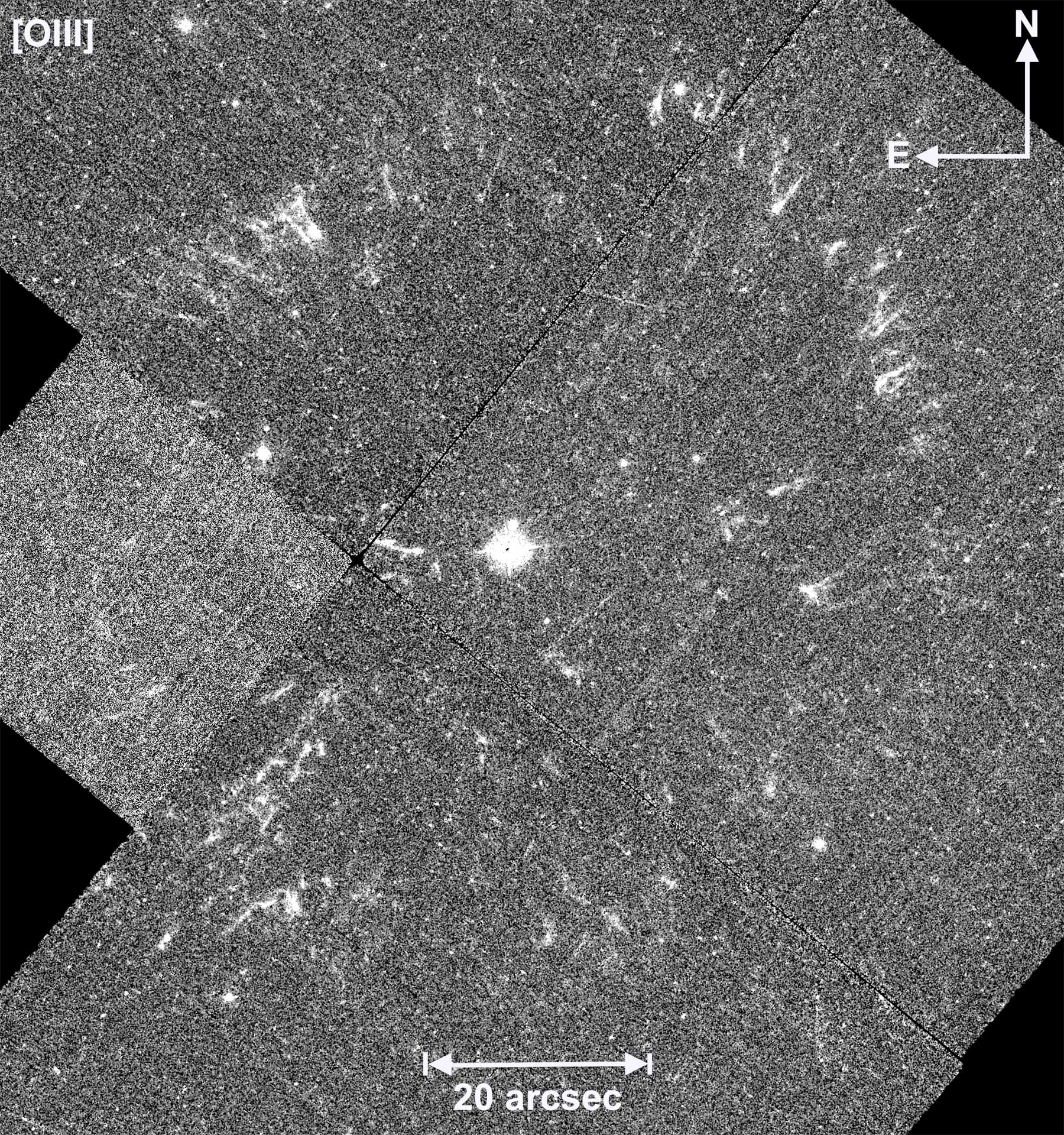}
\caption{HST/WFPC2 summed image in F502N [OIII] filter of the ejecta of the old nova GK Per}
\end{figure}

\clearpage

\begin{figure}
\figurenum{4}
\epsscale{1.0}
\plotone{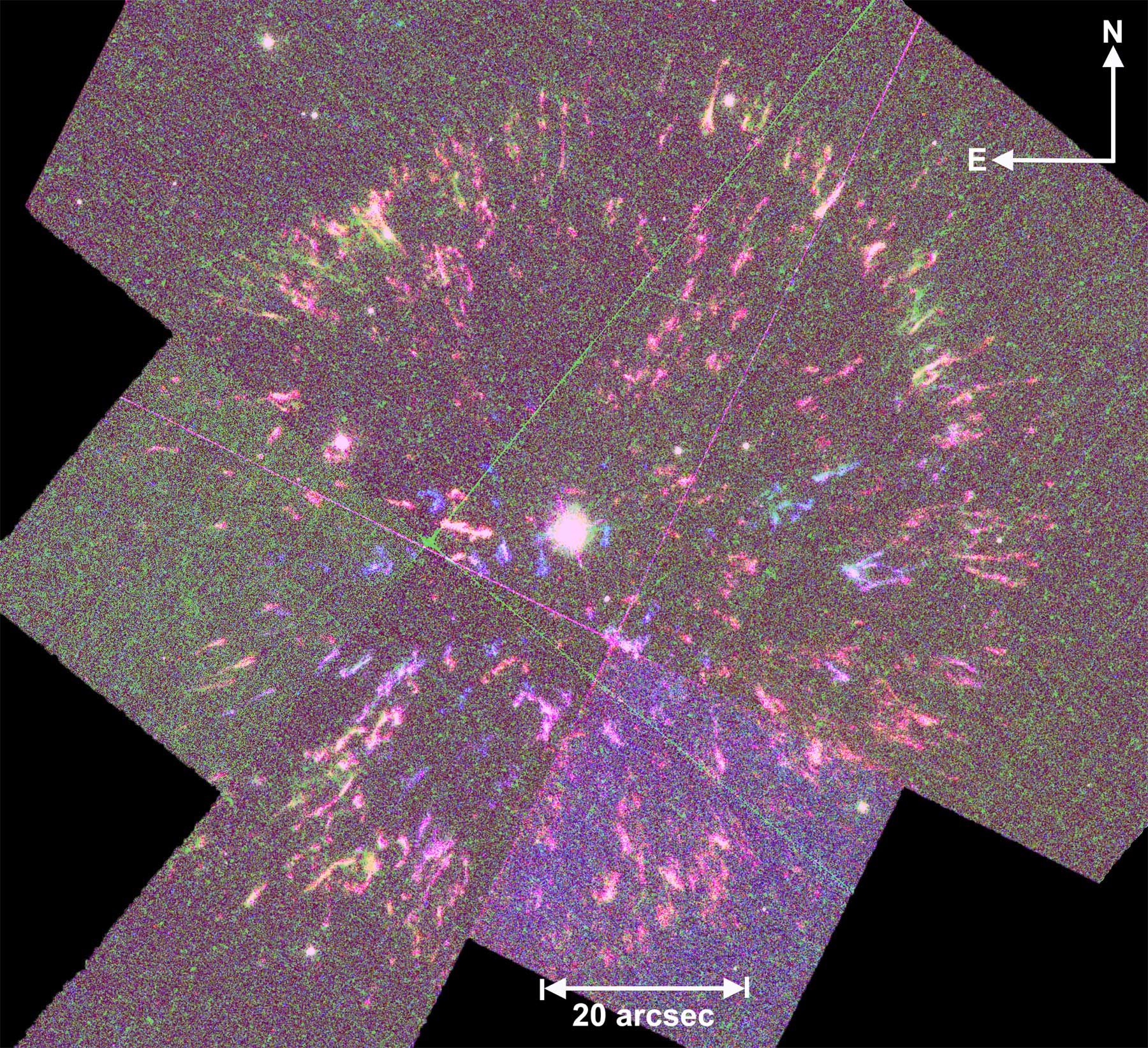}
\caption{The sum of Figures 1, 2 and 3 where red is [NII], green is [OIII] and blue is H$\alpha$. }
\end{figure}

\clearpage

\begin{figure}
\figurenum{5}
\epsscale{1.0}
\plotone{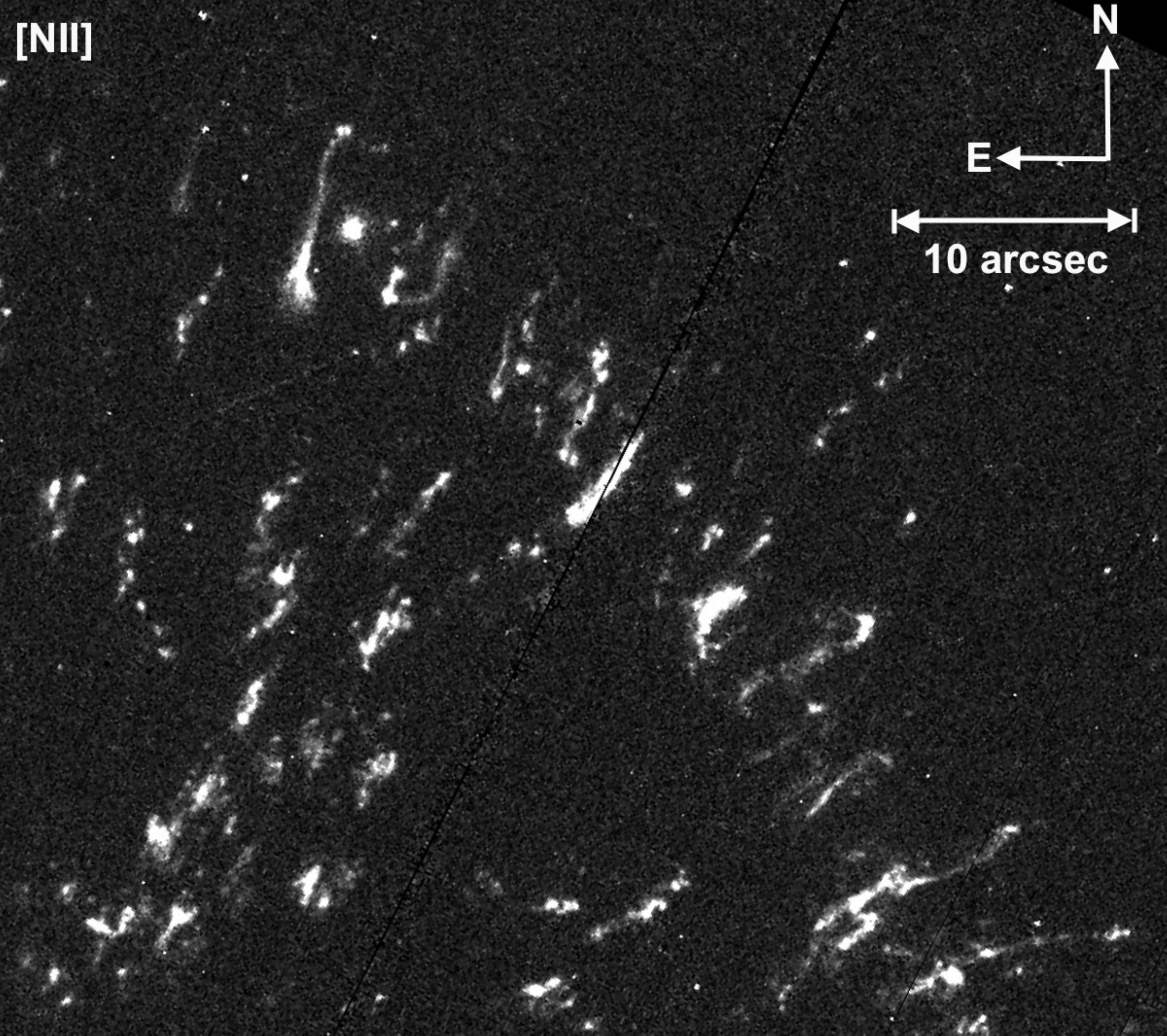}
\caption{A close-up of a section of the HST [NII] image of GK Per. The striking cometary-like structures are discussed in the text.}
\end{figure}

\clearpage

\begin{figure}
\figurenum{6}
\epsscale{0.8}
\plotone{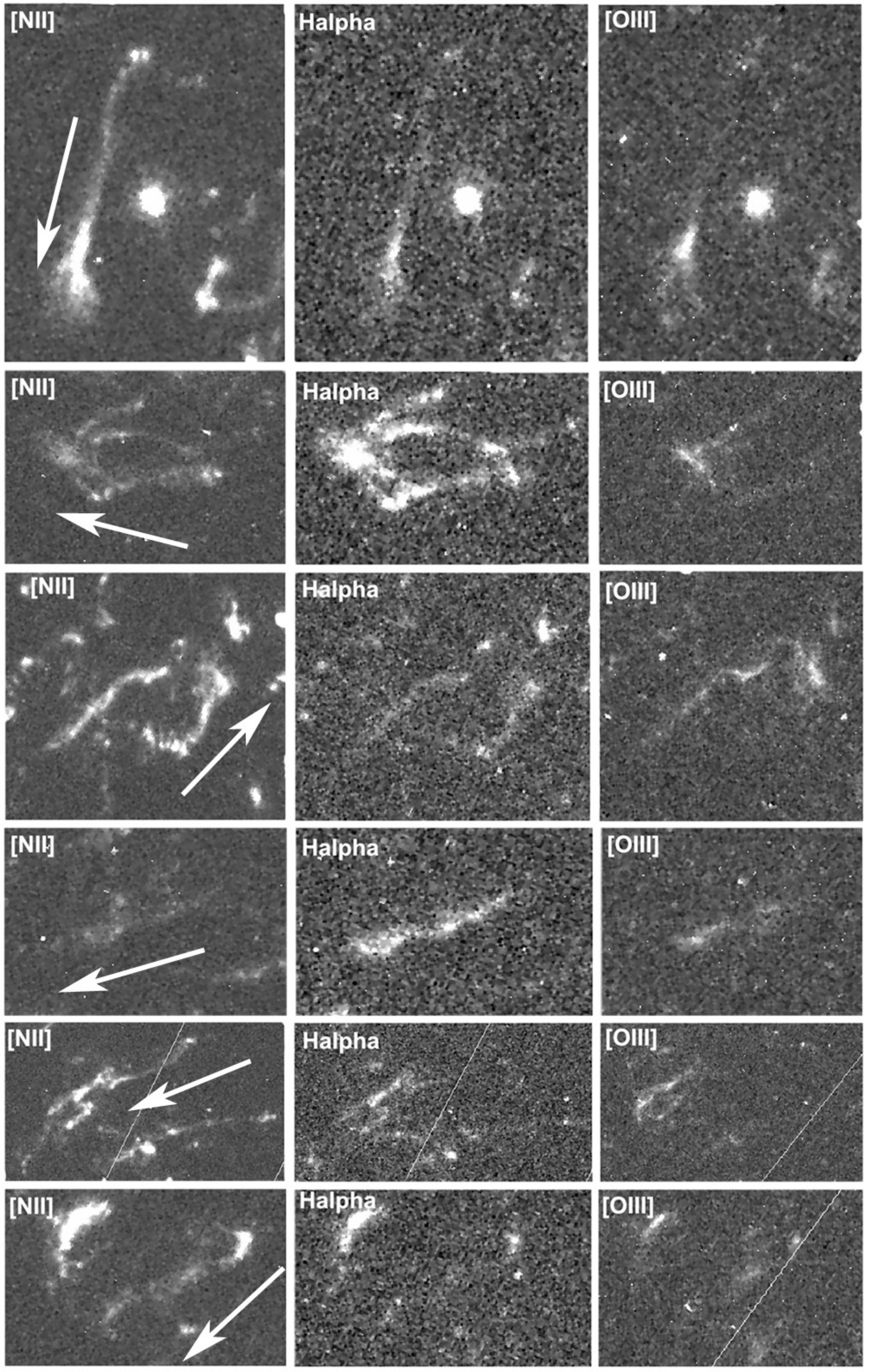}
\caption{A mosaic closeup (displaying three filters) of several of the cometary features in the GK Per ejecta imaged with HST/WFPC2. Arrows point back at GK Per itself.}
\end{figure}

\clearpage

\begin{figure}
\figurenum{7a}
\epsscale{1.0}
\plotone{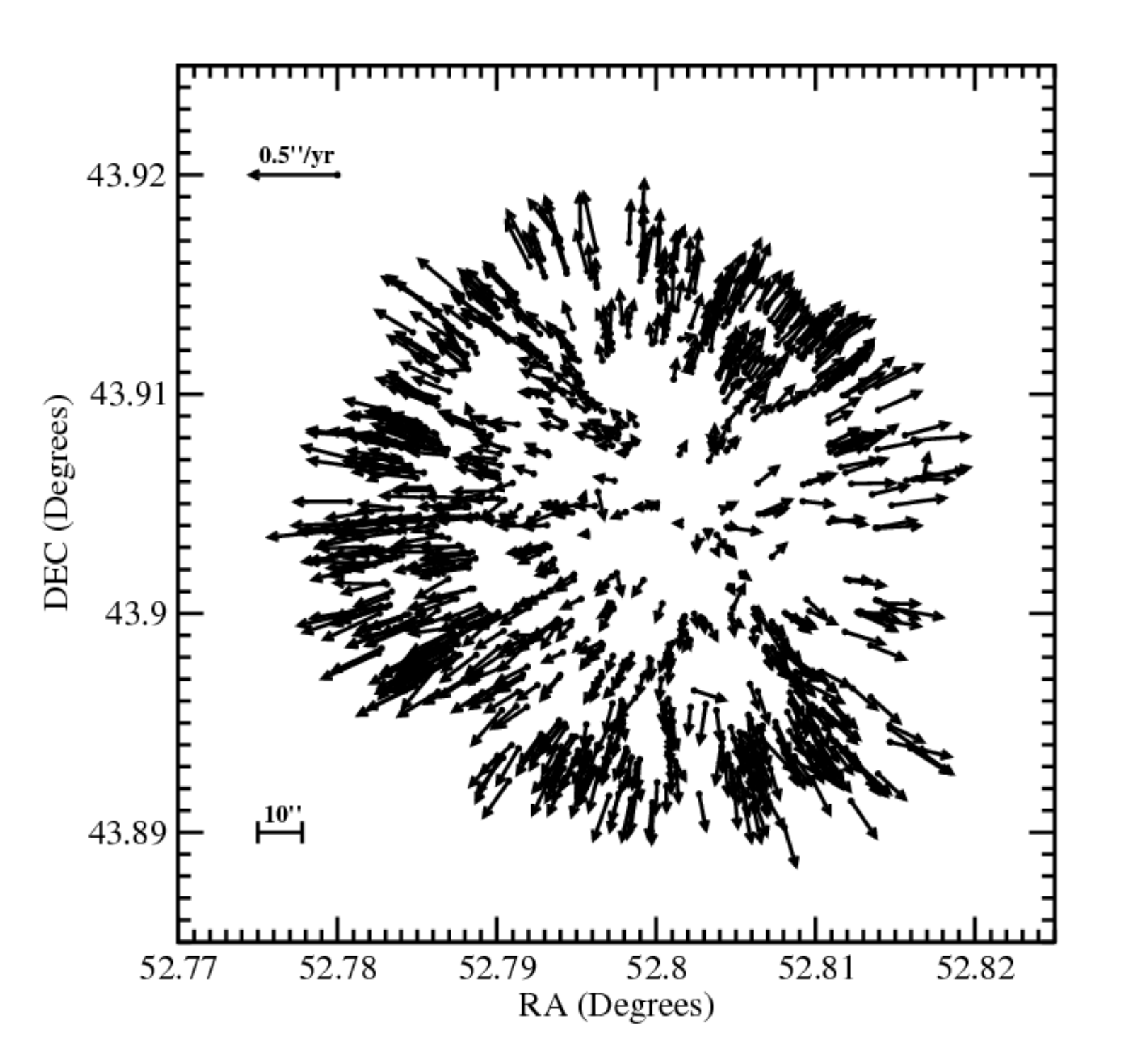}
\caption{The proper motions of 937 knots imaged from Figure 1. The vectors of the yearly motion have been scaled by a factor of 500 to make them more visible. We have placed in the upper left a scale vector of 0.5 arcsec/yr.}
\end{figure}

\clearpage

\begin{figure}
\figurenum{7b}
\epsscale{1.0}
\plotone{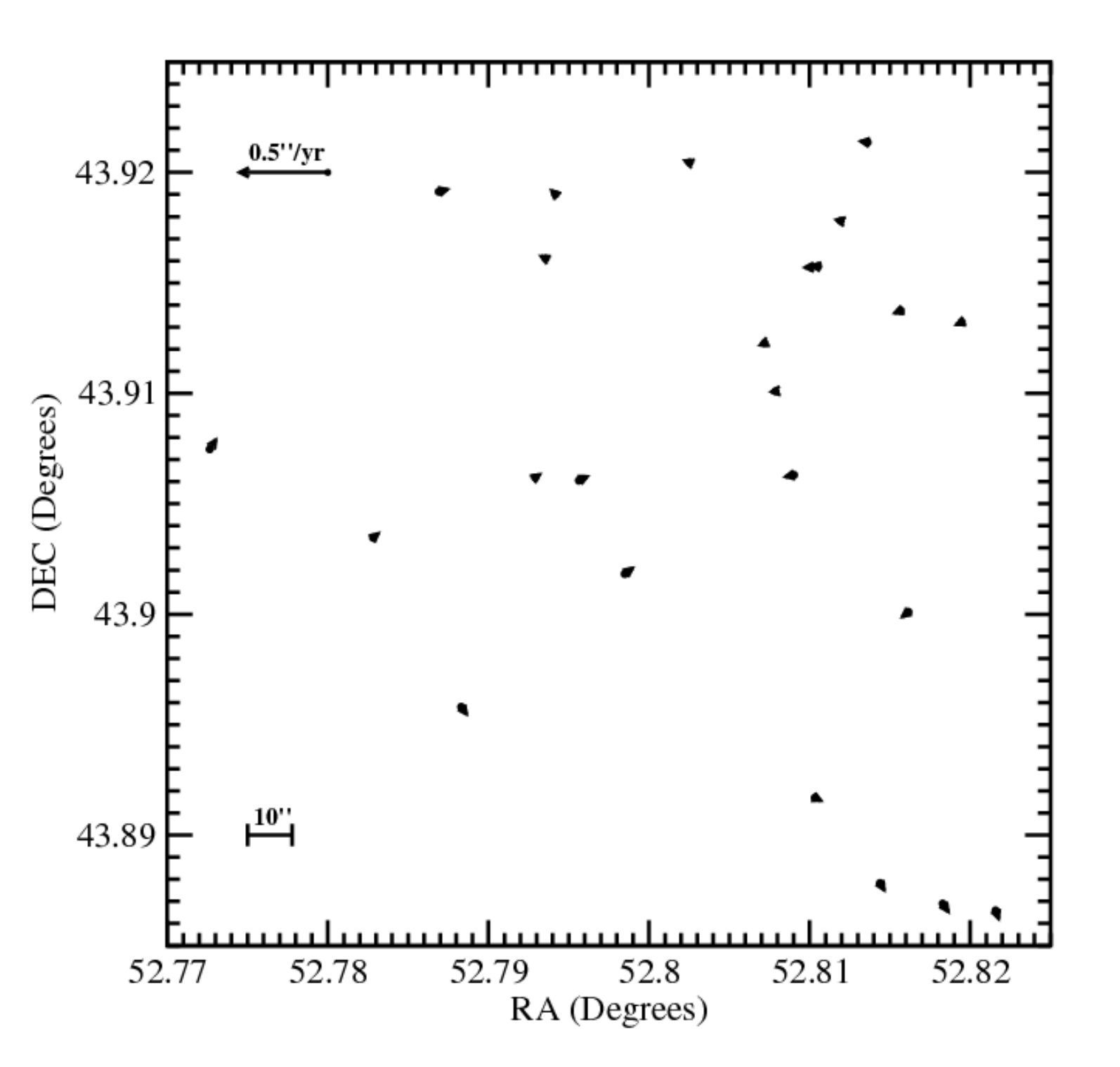}
\caption{The proper motions of 25 stars from Figure 1. The vectors of the measured yearly motion of the stars have been scaled by a factor of 500, the same scale factor as the knots. In all cases the measured motion is less than $\sim 0.04$ arcsec/yr.}
\end{figure}

\clearpage

\begin{figure}
\figurenum{8}
\epsscale{1.2}
\plotone{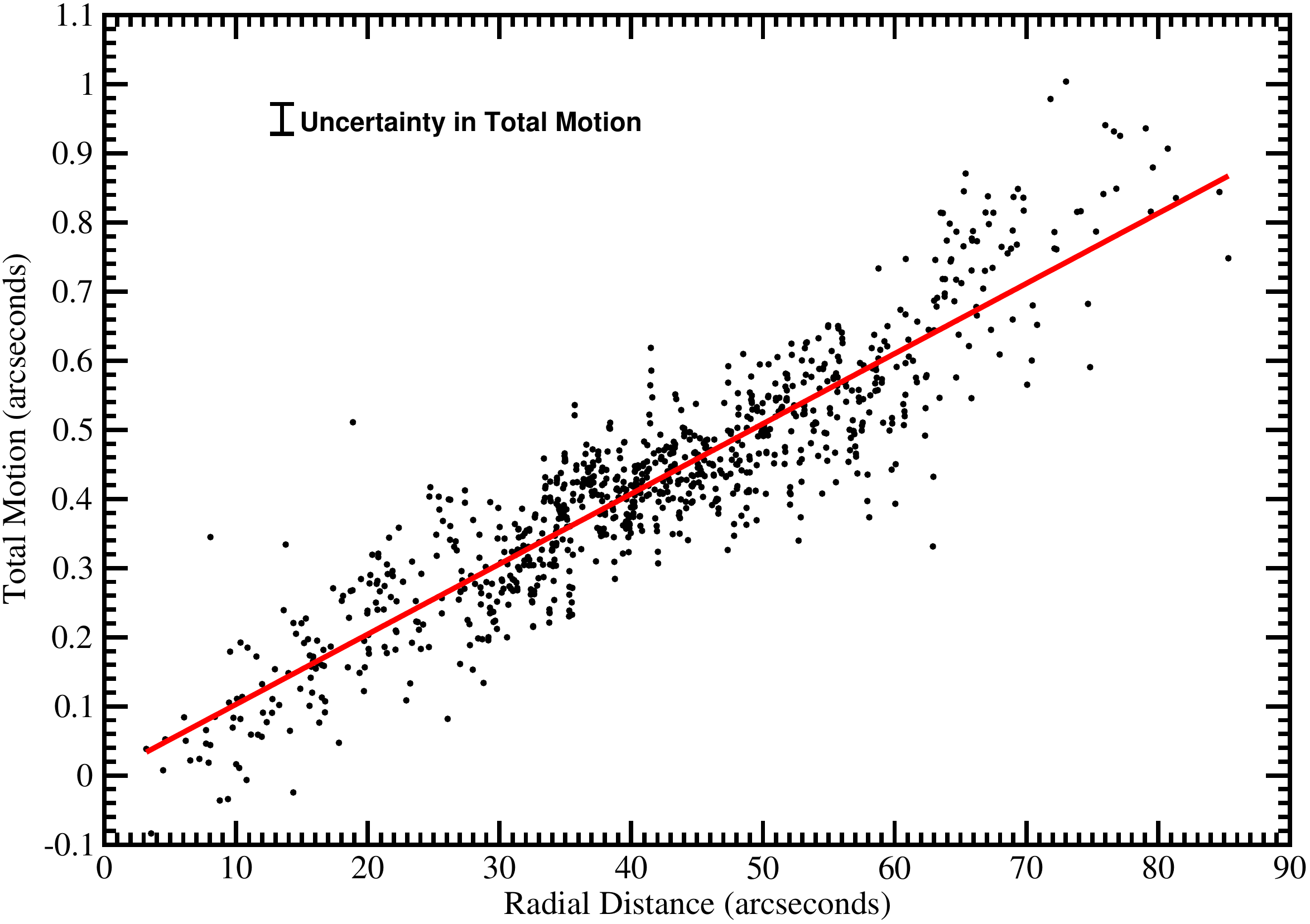}
\caption{The amplitude of the knots'  (of Figure 1) proper motions as a function of radial distance from GK Per. A typical error bar in the total motion (y-axis) for each data point is shown in the top left corner of the figure. A typical error bar in the x-axis is the size of each data point. A least squres linear fit to the data is shown. The vertical scatter in this Hubble flow is due to the non-isotropy of the nova ejection velocity field as seen in Figure 7a.}
\end{figure}

\clearpage

\begin{figure}
\figurenum{9}
\epsscale{0.9}
\plotone{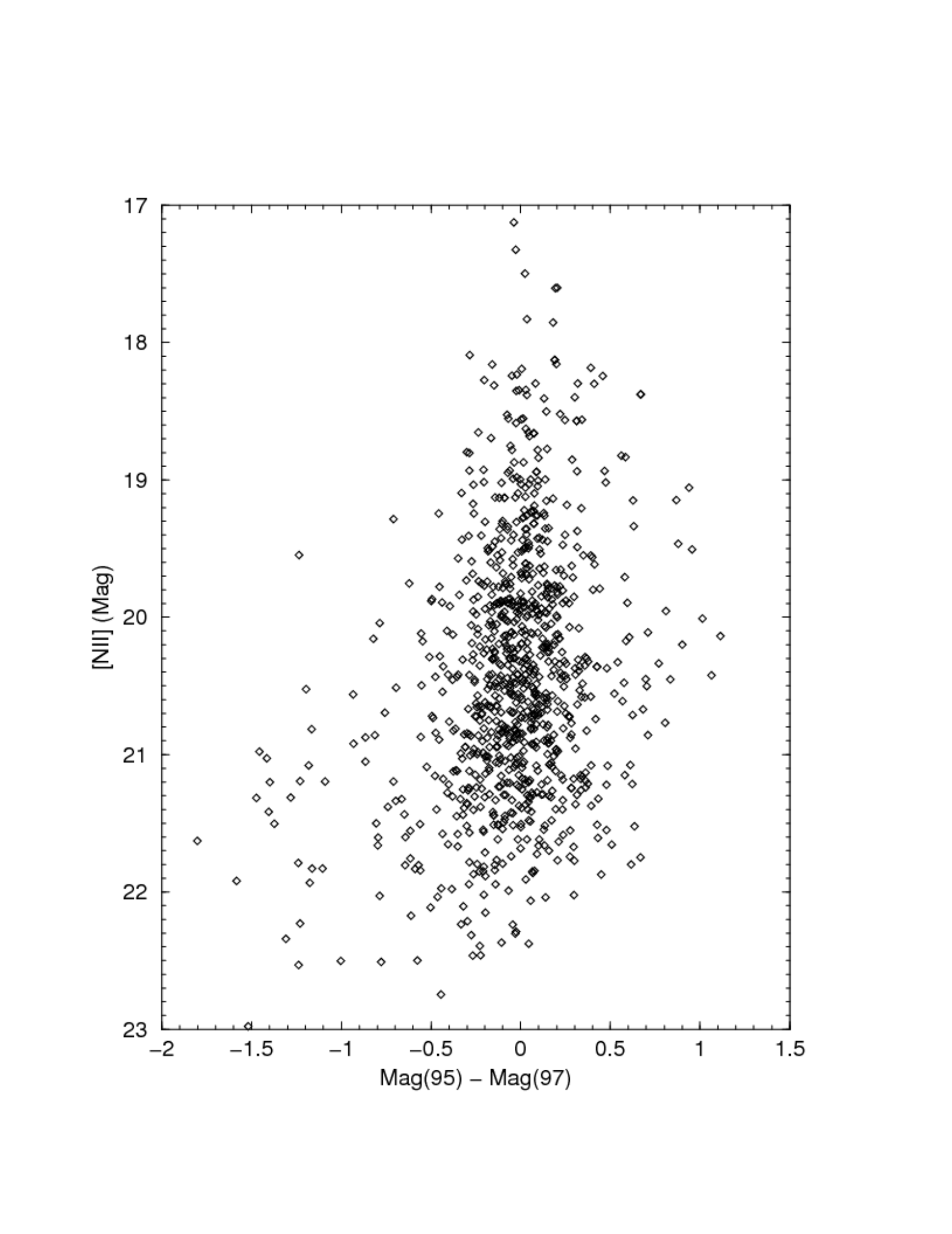}
\caption{A comparison between the brightnesses of knots in the [NII] images measured in October 1995 and January 1997. Knots on the left side of Figure 10 were brighter in January 1997 while the knots to the right of most points were brighter in October 1995. The error bars are comparable to the sizes of the data points.}
\end{figure}

\clearpage

\begin{figure}
\figurenum{10}
\epsscale{1.0}
\plotone{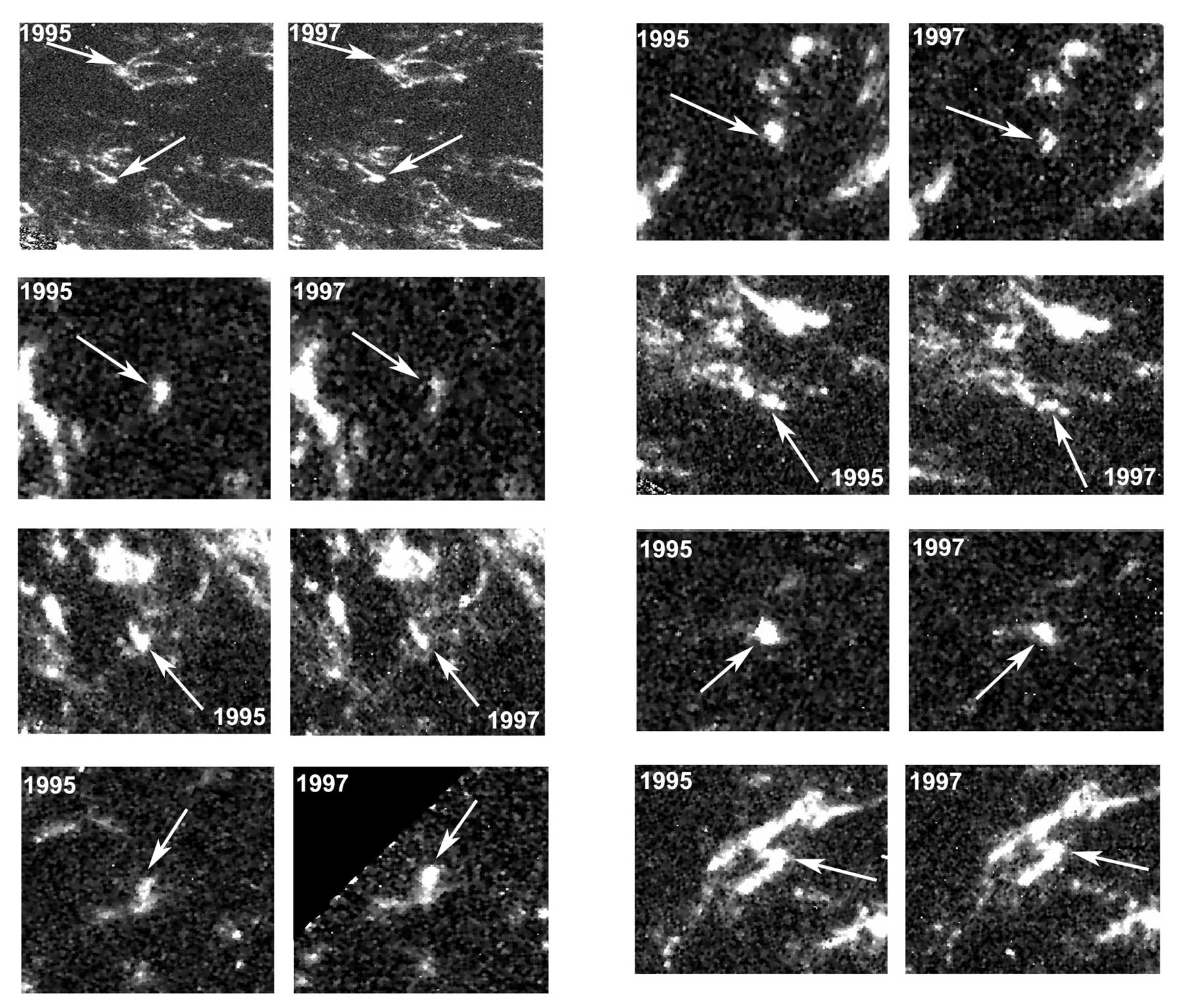}
\caption{Narrowband [NII] images of a few of the knots that changed most in brightness between October 1995 and January 1997.}
\end{figure}

\clearpage

\begin{figure}
\figurenum{11a}
\epsscale{0.6}
\plotone{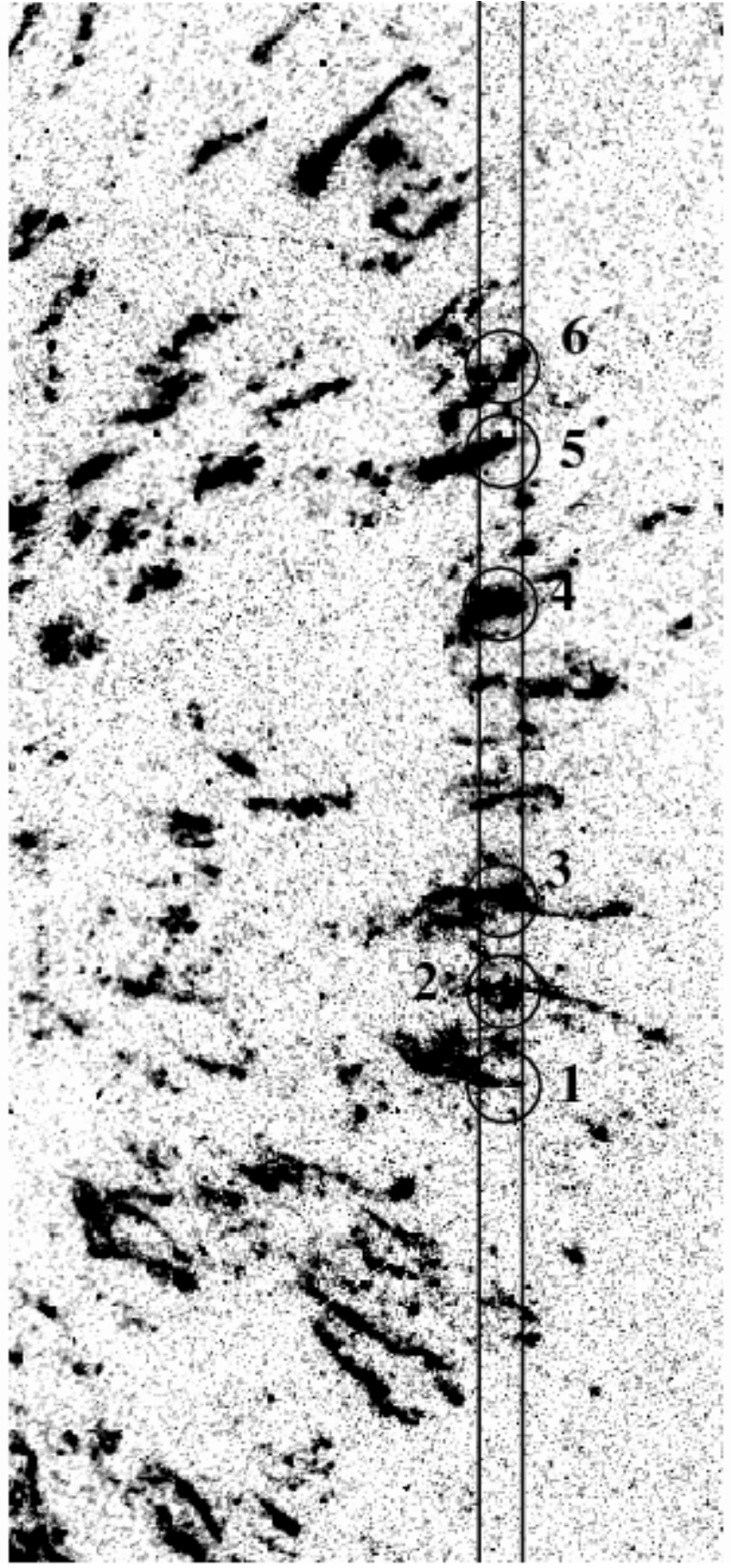}
\caption{Position of the slit on six bright knots in the ejecta of GK Per.}
\end{figure}

\clearpage

\begin{figure}
\figurenum{11b}
\epsscale{0.7}
\plotone{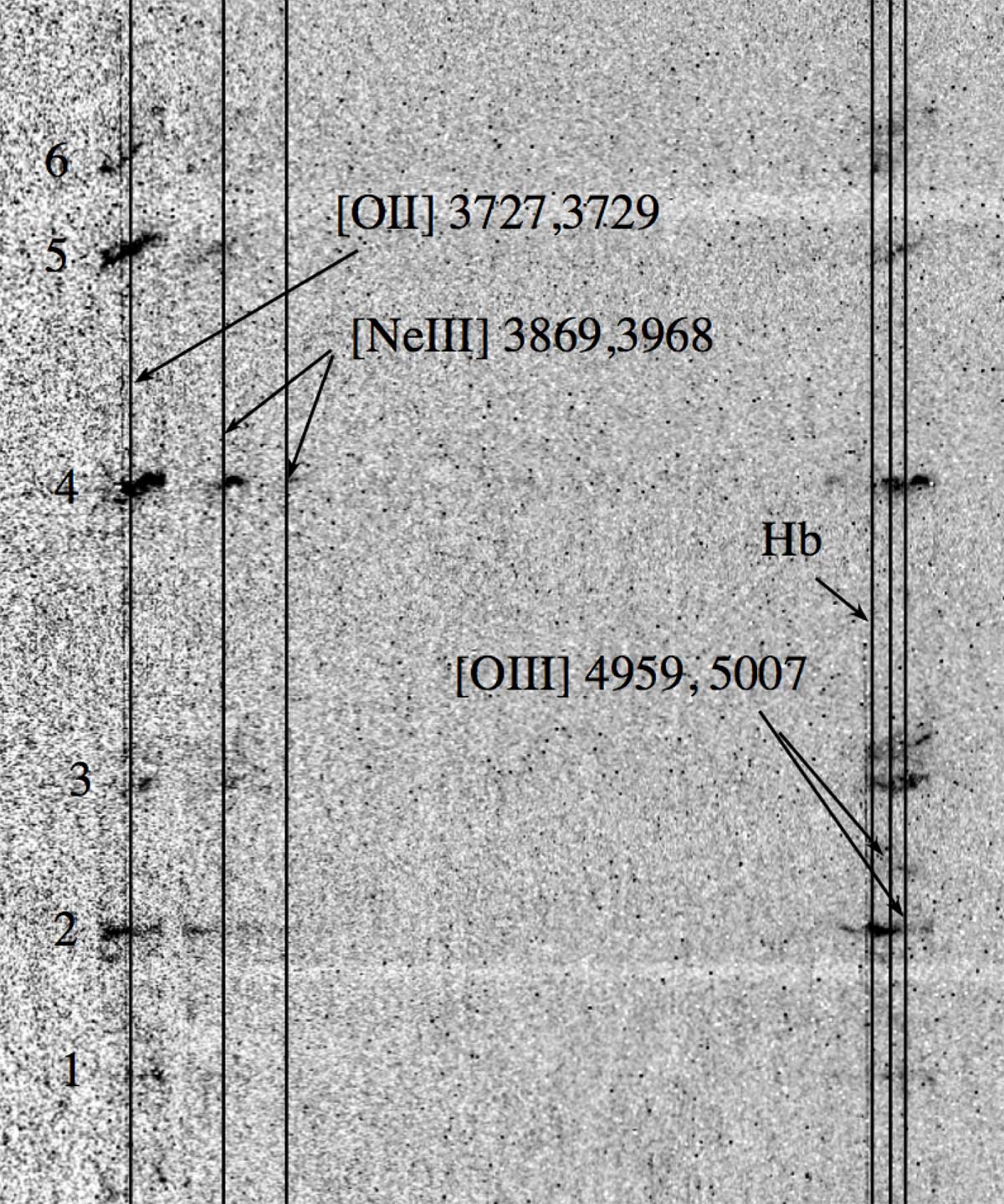}
\caption{The resulting spectra consist of very broad emission lines, but no continuum, as is expected of moderate-density nebular gas. The spectral region from [OII] through [OIII] is shown here.}
\end{figure}

\clearpage

\begin{figure}
\figurenum{11c}
\epsscale{0.6}
\plotone{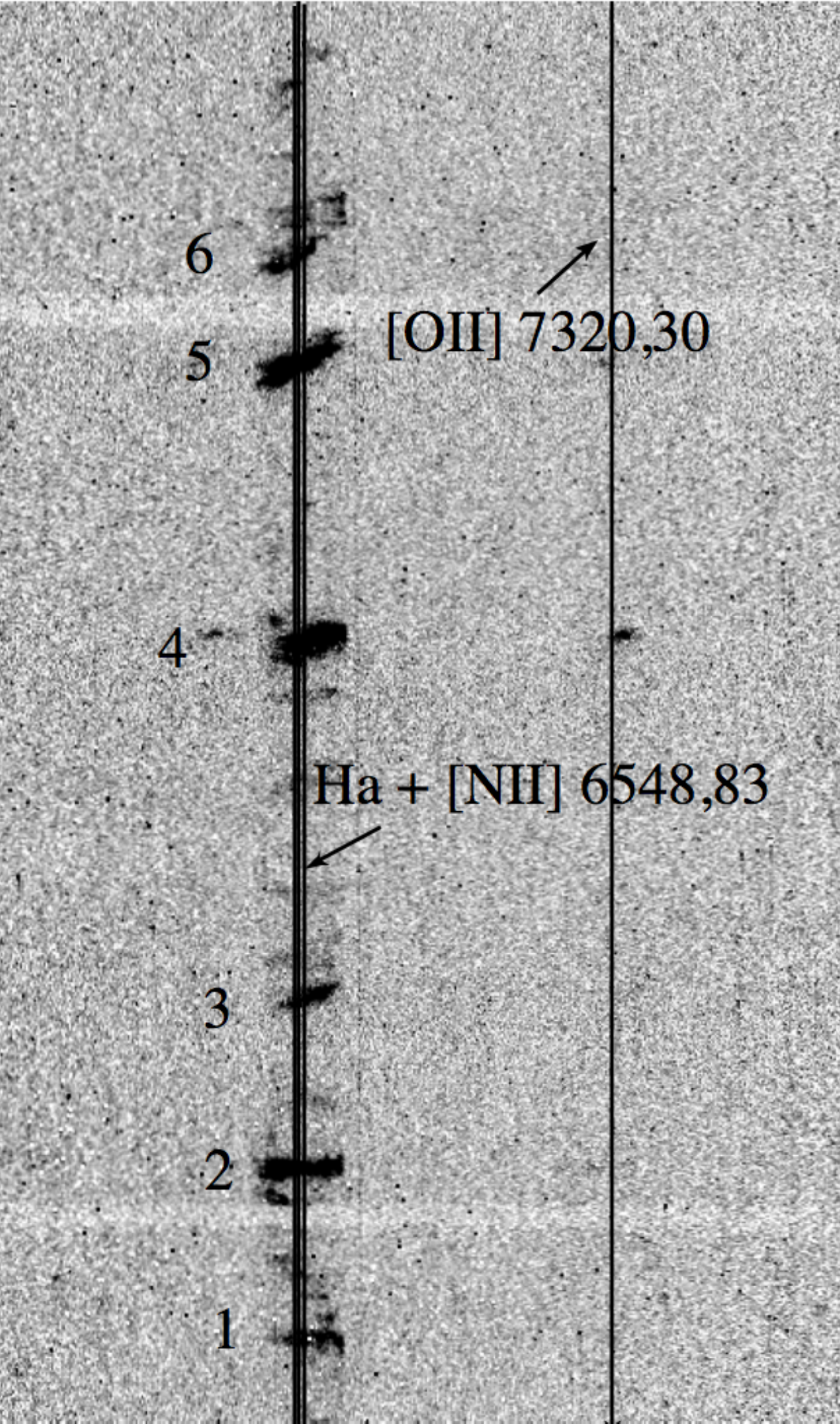}
\caption{Same as Figure 11b except that the spectral region shown covers the H$\alpha$ + [NII] lines.}
\end{figure}

\clearpage

\begin{figure}
\figurenum{12a}
\epsscale{1.0}
\plotone{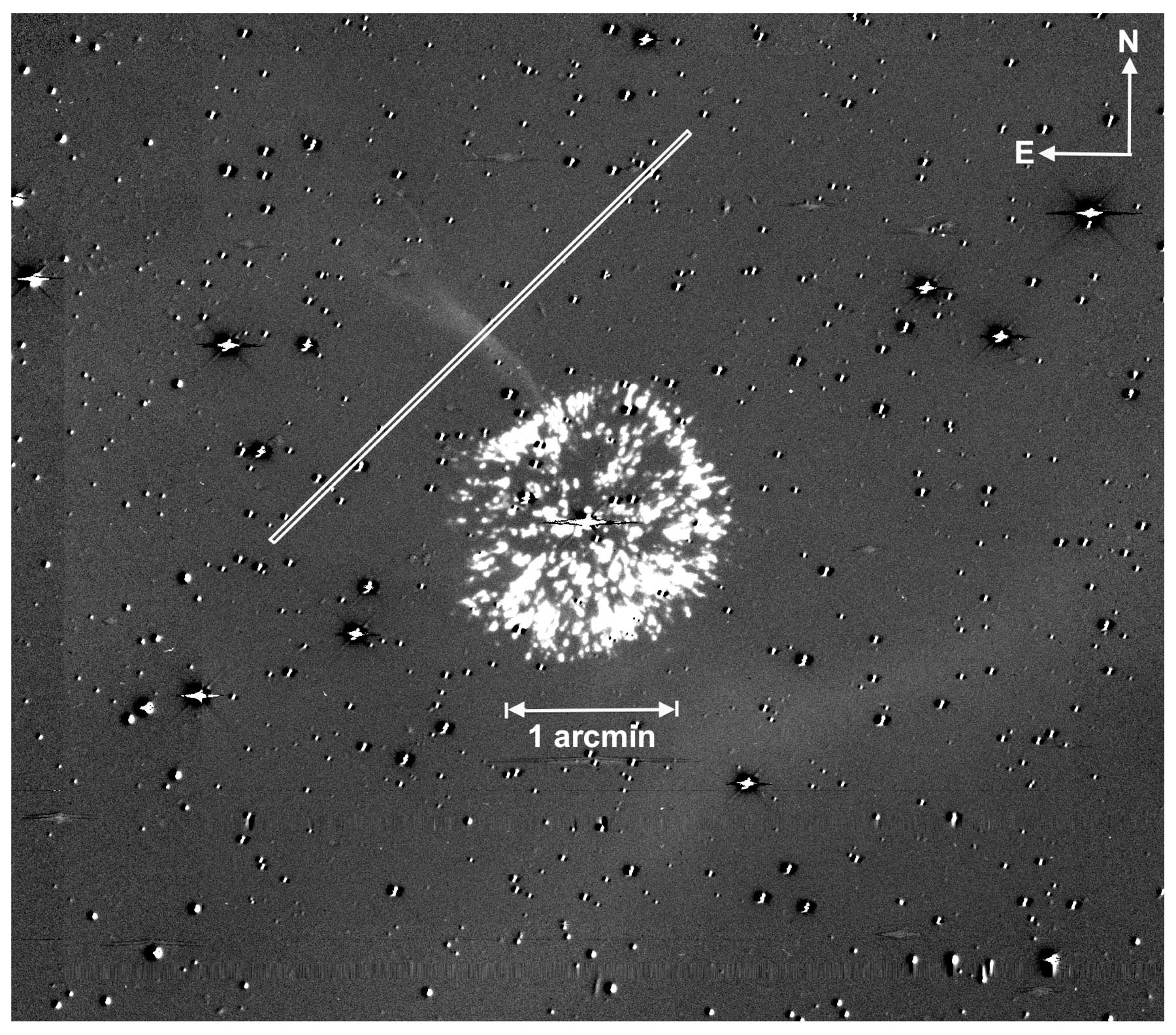}
\caption{A net (H$\alpha$ + [NII] minus broadband R) image of the ejecta of GK Per, showing the``jet-like" feature extending to the Northeast, and terminating in a very faint fan-like structure. The placement of the RC spectrograph slit is also shown in the image.}
\end{figure}

\clearpage

\begin{figure}
\figurenum{12b}
\epsscale{1.11}
\plotone{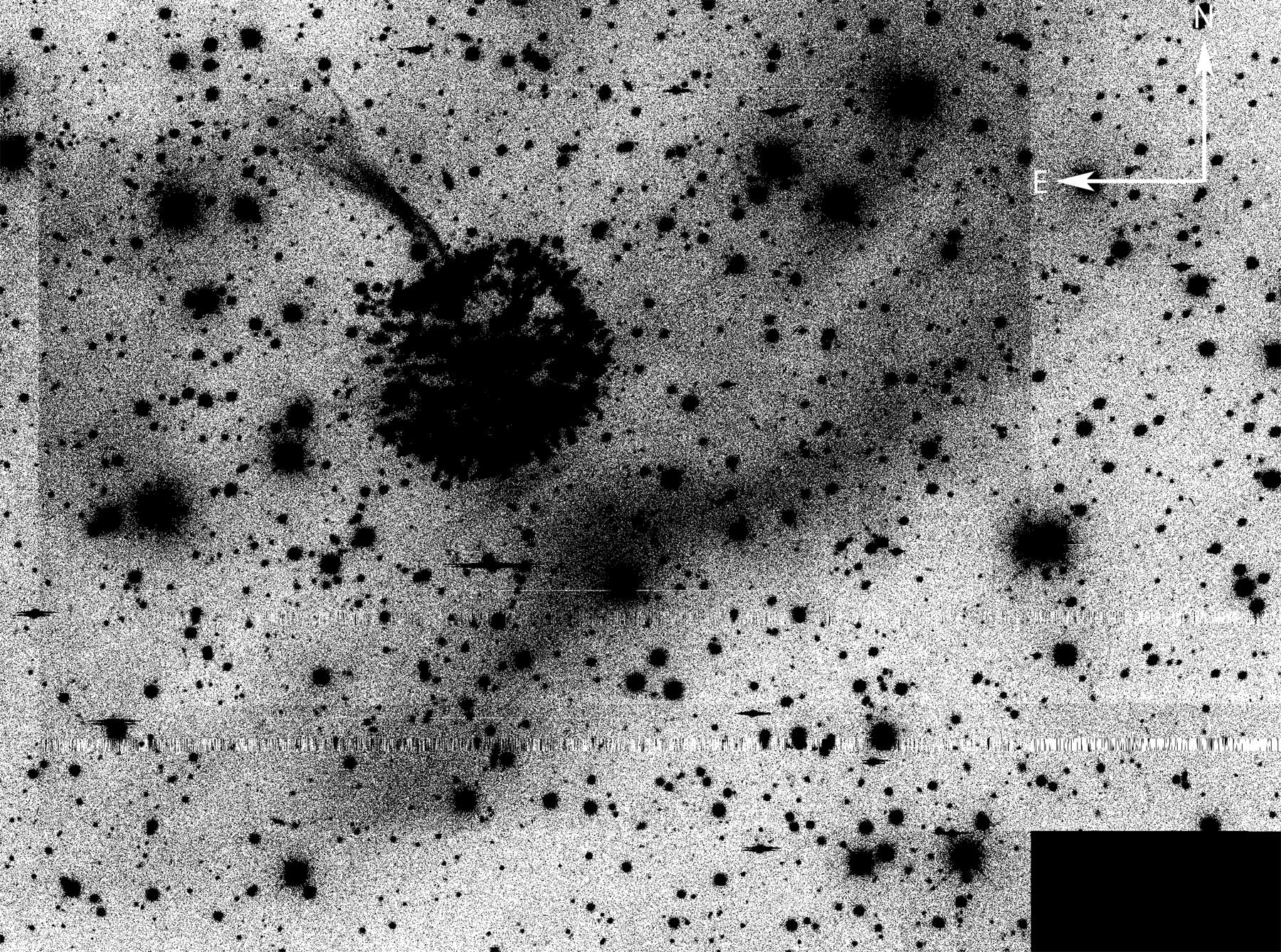}
\caption{Same as Figure 12a but stretched and shown as a negative to emphasize the faint but large-scale matter oriented NW to SE.}
\end{figure}

\clearpage

\begin{figure}
\figurenum{13}
\epsscale{0.55}
\plotone{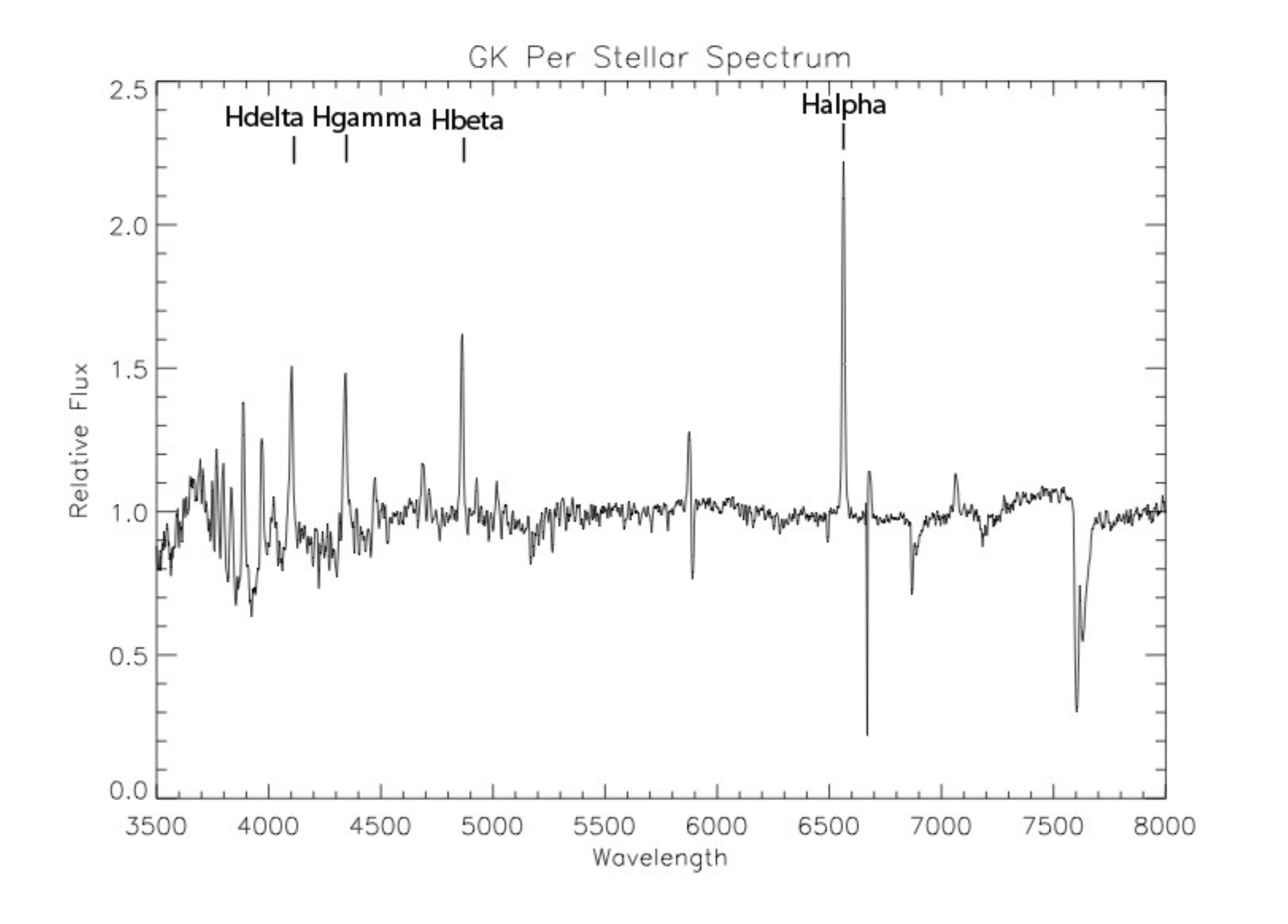}
\epsscale{0.55}
\plotone{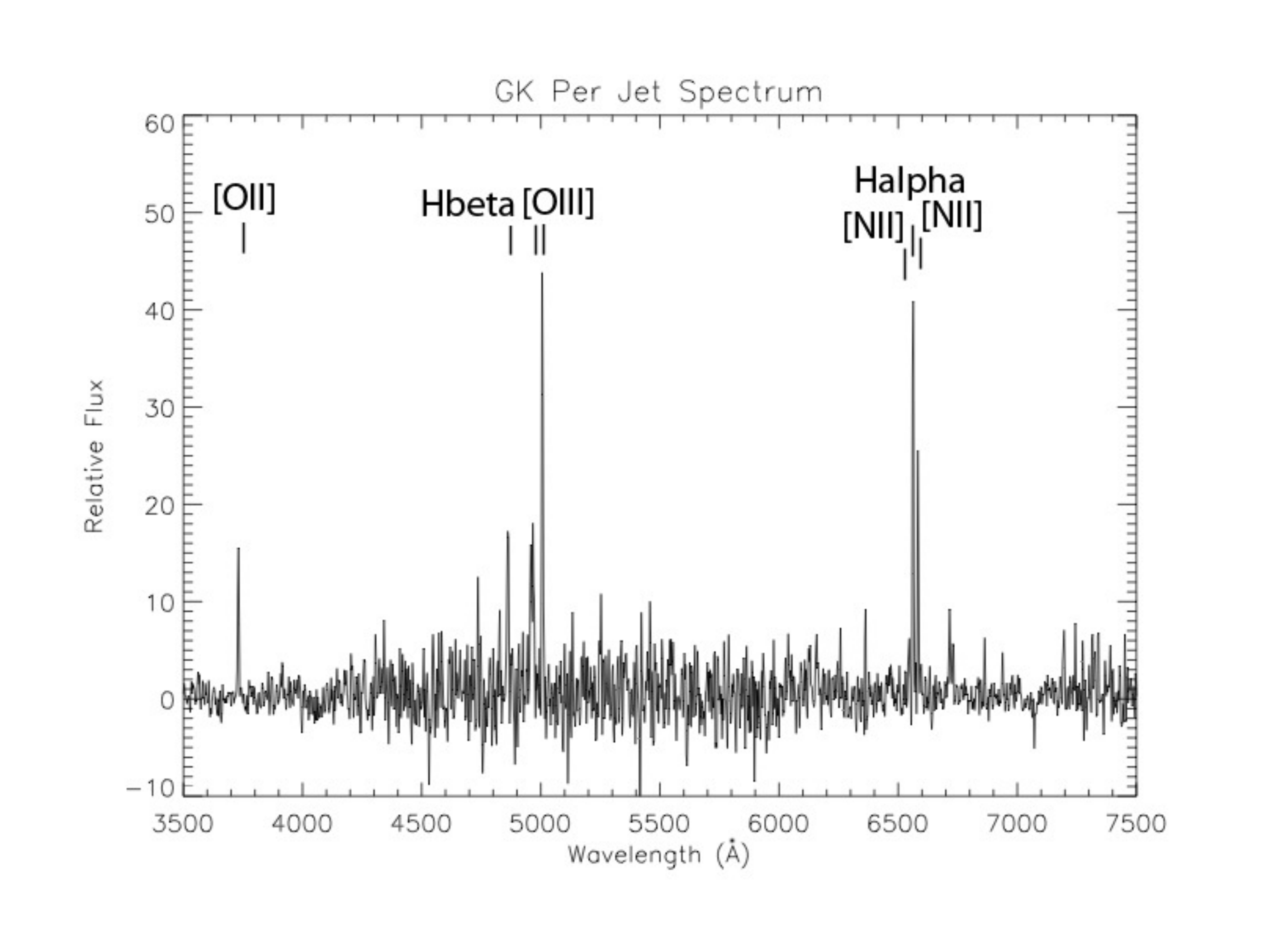}
\epsscale{0.55}
\plotone{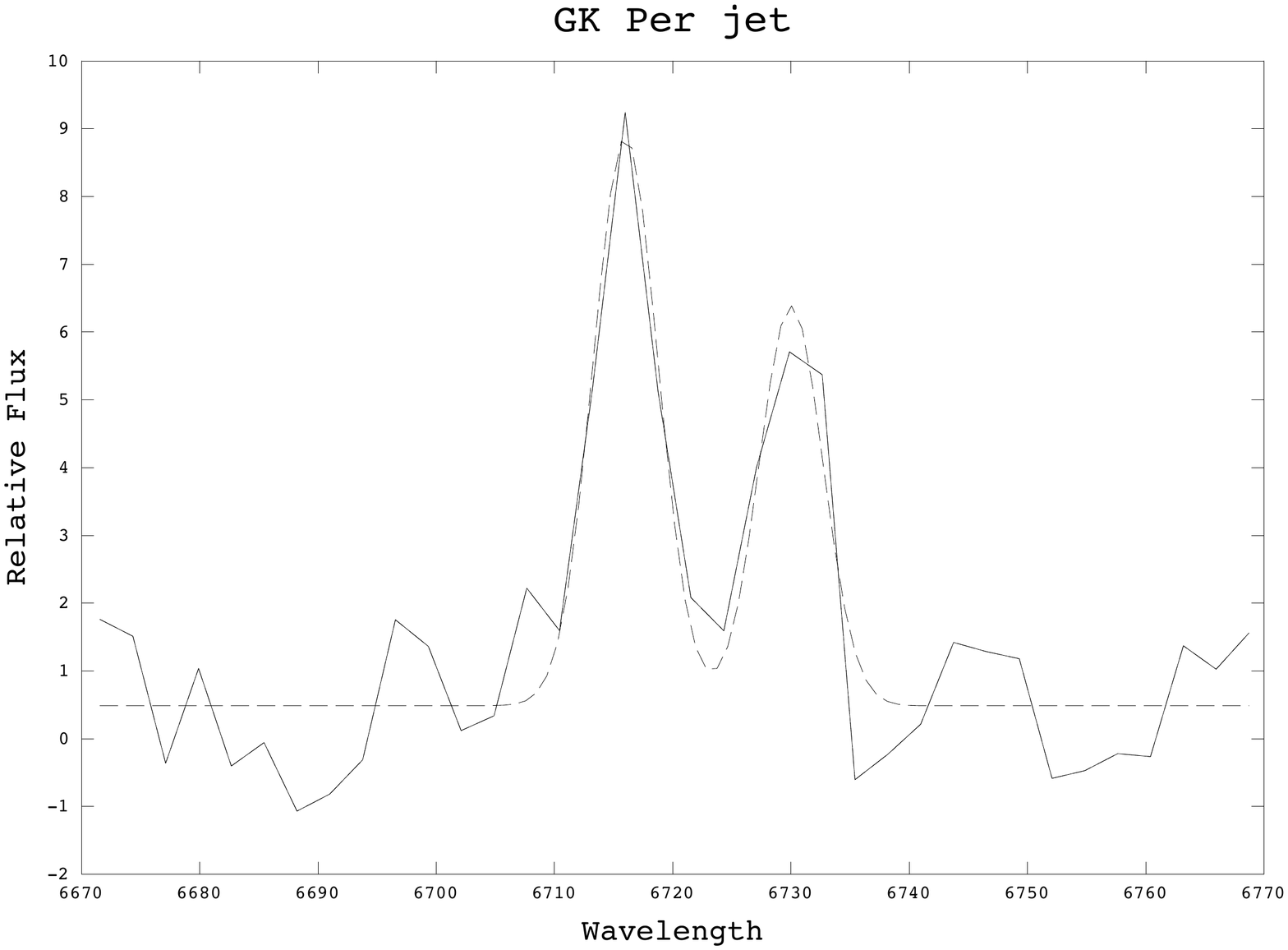} 
\caption{Top Panel: The spectrum of GK Per. Middle Panel: The spectrum of the GK Per ``jet" dominated by emission lines of H$\alpha$, [NII], [OIII], [OII] and [SII]. Bottom Panel: The [SII] doublet used to derive the density of matter in the "jet".} 
\end{figure}

\clearpage

\begin{deluxetable}{lccc}
\tabletypesize{\footnotesize}
\tablecaption{Photometry and Spectroscopy}
\tablewidth{0pt}
\tablehead{
\colhead{Date} & \colhead{Exposure Time} & \colhead{Filter} &  \colhead{Proposal ID} \\
DD-MM-YYYY & (sec) & &}
\startdata
\multicolumn{4}{c}{HST/ WFPC2} \\
\hline
08-11-1995 & 1400.0 & F658N & 6060 \\
08-11-1995 & 700.0 & F656N & 6060 \\
08-01-1997 & 1200.0 & F658N & 6965 \\
08-01-1997 & 180.0 & F675W & 6965 \\
08-01-1997 & 180.0 & F555W & 6965 \\
08-01-1997 & 180.0 & F450W & 6965 \\
08-01-1997 & 4600.0 & F469N & 6965 \\
08-01-1997 & 2600.0 & F487N & 6965 \\
08-01-1997 & 2600.0 & F502N & 6965 \\
09-01-1997 & 2600.0 & F375N & 6965 \\
09-01-1997 & 5200.0 & F343N & 6965 \\
\hline
\multicolumn{4}{c}{HST/STIS} \\
\hline
29-01-1999 & 4886.0 & G430L & 7768\tablenotemark{a} \\
29-01-1999 & 2586.0 & G750L & 7768\tablenotemark{a} \\
\hline
 \multicolumn{4}{c}{KPNO 4m/CCD Mosaic Imager} \\
\hline
06/07-02-2010 & 7440.0 & H$\alpha$ k1009 & 10A-0621 \\
06/07-02-2010 & 720.0 & Harris R k1004 & 10A-0621 \\
\hline
\multicolumn{4}{c}{KPNO 4m/RC Spectrograph} \\
\hline
10-02-2010 & 180.0 & t2kb & 10A-0621\tablenotemark{b} \\
10-02-2010 & 9600.0 & t2kb & 10A-0621\tablenotemark{c} \\
\enddata
\tablenotetext{a}{GK Per spectrum of the knots}
\tablenotetext{b}{GK Per spectrum of the central star}
\tablenotetext{c}{GK Per spectra of the jet}
\end{deluxetable}

\clearpage

\begin{table}
\begin{center}
\caption{Observed emission line fluxes (10$^{-14}$erg~cm$^{-2}$~s${-1}$) for the six brightest knots. These were measured on the extracted spectra and multiplied by a diffuse-to-point factor of 0.114 and 0.131~arcsec$^2$, for the G430L and G750L gratings respectively.}
\begin{tabular}{cccccccc}
\hline
Knot  & [O~{\sc ii}]& He~{\sc i} & [Ne~{\sc iii}] &H$\beta$& [O~{\sc iii}] & [N~{\sc ii}]-H$\alpha$ & [O~{\sc ii}] \\
      & $\lambda \lambda$3727,29 & $\lambda$3889 & $\lambda$3969 & & $\lambda \lambda$4959,5007 & $\lambda \lambda$6548,83-6561 & $\lambda \lambda$7320,30 \\
\hline
1    &  1.53& --  & --  &   --     & --  &       4.52&  --\\
2    &  2.37& 1.60& --  &   --     & 2.80&       9.60&  --\\
3    &  1.15& 0.63& --  &   --     & 2.26&       3.86&  --\\
4    &  7.25& 1.97& 1.07&   0.25   & 1.70&       22.9&  1.56\\
5    &  4.34& 1.77& --  &   --     & 1.00&       12.7&  0.73\\
6    &  1.67& --  & --  &   --     & 0.95&       2.69&  --\\
\hline
\end{tabular}
\end{center}
\end{table}

\clearpage

\begin{table}
\begin{center}
\caption{Dereddened ($E(B-V)=0.32$~mag) emission line flux ratios for the six brightest knots.}
\begin{tabular}{ccccccc}
\hline
Knot &
[O~{\sc ii}]$\lambda$3727/ &He~{\sc i}$\lambda$3889/ &[Ne~{\sc iii}]$\lambda$3969/& [O~{\sc iii}]/ & [N~{\sc ii}]$^2$-H$\alpha$/ &[O~{\sc ii}]$\lambda \lambda$7320,30/\\
&
[O~{\sc iii}]$^1$          &[O~{\sc iii}]            &H$\beta$                    & H$\beta$       & [O~{\sc iii}]               &[O~{\sc ii}]$\lambda \lambda$3727,29 \\
1&    --       &     --       &           -- &     --       &     --       &     --       \\
2&    1.23     &     0.80     &           -- &     --       &     2.97     &     --       \\
3&    0.74     &     0.39     &           -- &     --       &     1.48     &     --       \\
4&    6.19     &     1.62     &     5.71     &     6.64     &     11.6     &     0.11     \\
5&    6.30     &     2.47     &           -- &     --       &     11.0     &     0.09     \\
6&    1.24     &     --       &           -- &     --       &     1.19     &     --       \\
\hline
\multicolumn{7}{l}{$^1$[O~{\sc iii}] = [O~{\sc iii}] $\lambda \lambda$4959,5007.}\\
\multicolumn{7}{l}{$^2$[N~{\sc ii}] = [N~{\sc ii}] $\lambda \lambda$6548,83.  }\\
\end{tabular}
\end{center}
\end{table}

\clearpage

\begin{table}
\begin{center}
\caption{Jet Emission line strengths relative to [O~{\sc iii}].}
\begin{tabular}{ccc}
\hline
Emission Line & Wavelength & Relative Strength \\
\hline
[O~{\sc ii}] & 3729 & 0.58 \\

H$\beta$ & 4861 & 0.60 \\

[O~{\sc iii}] & 5007 & 1.00 \\

[N~{\sc ii}] & 6545 & 0.17 \\

H$\alpha$ & 6562 & 0.99 \\

[N~{\sc ii}] & 6583 & 0.54 \\

[S~{\sc ii}] & 6716 & 0.73 \\

[S~{\sc ii}] & 6731 & 0.52 \\

\hline
\end{tabular}
\end{center}
\end{table}

\end{document}